\documentclass[journal,doublecolumn]{IEEEtran}

\IEEEoverridecommandlockouts
\usepackage{cite}
\usepackage{amsthm,amsmath,amssymb}
\usepackage{algorithm}
\usepackage{algorithmic}
\usepackage{setspace}
\usepackage{graphicx}
\usepackage{textcomp}
\usepackage{xcolor}
\usepackage{stfloats}
\usepackage{color}
\usepackage{multirow}
\usepackage{array}
\usepackage{bm}
\usepackage{subfig}
\usepackage{booktabs}
\usepackage{lettrine}
\usepackage[nolist]{acronym}
\usepackage{subcaption}

\usepackage{caption}
\newtheorem{Remark}{Remark}

\usepackage{authblk}
\usepackage[implicit=true]{hyperref} 
\hypersetup{
	colorlinks=true,
	linkcolor=blue,
	citecolor=blue,
	bookmarksnumbered=false,
	bookmarks=false
}

\begin{document}

\title{ODMA-based MIMO Massive Unsourced Random Access with Soft-Output Polar Codes \\}
\author{\IEEEauthorblockN{Tianya Li, Xiaoran Zhang, Nan Hu, Yongpeng Wu, \textit{Senior Member, IEEE}, Wenjun Zhang, \textit{Fellow, IEEE}, Xiang-Gen Xia, \textit{Fellow, IEEE}, and Chengshan Xiao, \textit{Fellow, IEEE}}
	\thanks{{
	
			T. Li was with the Department of Electronic Engineering at Shanghai Jiao Tong University, Shanghai 200240, China. He is now with the Department of Wireless and Device Technology Research, China Mobile Research Institute, Beijing 100053, China. E-mail: litianya@chinamobile.com.

			X. Zhang, and N. Hu are with the Department of Wireless and Device Technology Research, China Mobile Research Institute, Beijing 100053, China. E-mails: \{zhangxiaoran, hunan\}@chinamobile.com.

			Y. Wu (Corresponding author), and W. Zhang are with the Department of Electronic Engineering at Shanghai Jiao Tong University, Shanghai 200240, China.
			E-mails: \{yongpeng.wu, zhangwenjun\}@sjtu.edu.cn.
			
			X.-G. Xia is with the Department of Electrical and Computer Engineering, University of Delaware, Newark, DE 19716, USA. E-mail: xianggen@udel.edu.

			C. Xiao is with the Department of Electrical and Computer Engineering, Lehigh University, Bethlehem, PA 18015, USA. E-mail: xiaoc@lehigh.edu.
		}
	}
}

\maketitle
\thispagestyle{empty}

\begin{acronym}
	\acro{5G}{fifth-generation}
	\acro{6G}{sixth-generation}
	\acro{IoT}{Internet-of-Things}
	\acro{umMTC}{ultra-massive machine-type communications}
	\acro{URA}{unsourced random access}
	\acro{UMA}{unsourced multiple access}
	\acro{GMAC}{Gaussian multiple access channel}
	\acro{TIN}{treating interference as noise}
	\acro{MIMO}{multiple-input multiple-output}
	\acro{CS}{compressed sensing}
	\acro{LDPC}{low-density parity-check}
	\acro{SIC}{successive interference cancellation}
	\acro{SPARC}{sparse regression code}
	\acro{AMP}{approximate message passing}
	\acro{IDMA}{interleave-division multiple access}
	\acro{ODMA}{on-off division multiple access}
	\acro{SC}{successive cancellation}
	\acro{SCL}{\ac{SC} list}
	\acro{CRC}{cyclic redundancy check}
	\acro{BP}{belief propagation}
	\acro{MAP}{\textit{Maximum a Posteriori}}
	\acro{HO}{hard-output}
	\acro{SO}{soft-output}
	\acro{NOPICE}{noisy pilot channel estimation}
	\acro{OMP}{orthogonal matching pursuit}
	\acro{APP}{\textit{a posteriori probability}}
	\acro{MMV}{multiple measurement vector}
	\acro{LLR}{log-likelihood ratio}
	\acro{BS}{base station}
	\acro{i.i.d.}{independent and identically distributed}
	\acro{AWGN}{additive white Gaussian noise}
	\acro{PUPE}{per-user probability of error}
	\acro{JADCE}{joint activity detection and channel estimation }
	\acro{MMSE}{minimum mean-squared error}
	\acro{LMMSE}{linear \ac{MMSE}}
	\acro{MP}{message-passing}
	\acro{MPA}{\ac{MP} algorithm}
	\acro{SN}{sum node}
	\acro{VN}{variable node}
	\acro{JDD}{joint detection and decoding}
\end{acronym}

\begin{abstract}
	
	\par This paper investigates the design of the on-off division multiple access (ODMA) transmission scheme for multiple-input multiple-output (MIMO) massive unsourced random access (URA) systems with soft-output (SO) polar codes.  First, a three-segment pilot-uncoupled coding scheme is introduced under the ODMA framework, which reduces the coding rate of the data segment without increasing the transmission overhead, improving the overall system performance. Building upon this architecture, a hierarchical pattern detection framework is developed. Specifically, a coarse-grained candidate set of transmission patterns is first identified through correlation operations. Based on this, a message-passing (MP)-based pattern detection algorithm is developed to iteratively estimate the posterior probabilities of transmission patterns, followed by the \textit{maximum a posteriori} (MAP) estimation to obtain the precise pattern detection result. Furthermore, a joint pattern detection and data decoding algorithm based on the bit-wise SO information of polar decoder is investigated, where the posterior probability information provided by the polar decoder is exploited to refine the pattern detection and contribute to an improved accuracy. In addition, by leveraging bit-wise SO information of the successive cancellation list polar decoder, an MP-based iterative decoding algorithm is developed to significantly enhance the decoding performance. The proposed scheme simultaneously exploits the transmission gain of uncoupled-ODMA framework, the coding gain of polar codes in the short-blocklength regime, and the iterative decoding gain enabled by SO information, while the computational complexity is significantly reduced through the hierarchical detection framework. Simulation results demonstrate that the proposed scheme achieves strong robustness against multi-user interference and provides more than $2$ dB performance gain in the PUPE regime of $10^{-3}$ over existing URA schemes in large-scale MIMO systems.
	
\end{abstract}


\section{Introduction} \label{sec-1}

\par With the accelerated advancement of digitization in society, the number of \ac{IoT} devices will experience an explosive growth to the scale of hundreds of billions in the next ten years\cite{wu2020wcm}. According to the vision and paradigm of IMT-2030, \ac{umMTC}, building upon the foundation of mMTC from the \ac{5G} era, has emerged as a pivotal communication scenario envisioned for realization in \ac{6G} mobile networks\cite{You2021SCIS}. This paradigm shift presents unprecedented challenges to the massive connectivity, system capacity, transmission efficiency, and energy efficiency of mobile communication networks. To effectively address these issues, \ac{URA}, first introduced by Polyanskiy in \cite{yury2017isit}, handles massive uncoordinated users through a shared common codebook to reduce signaling overhead and improve efficiency,  exhibiting strong potential in massive random access scenarios.

\par With the emergence of the promising \ac{URA} paradigm, substantial research efforts have been devoted to exploring low-complexity, high-spectral-efficiency, and high-performance coding schemes. These studies can be broadly categorized into several classes including slotted ALOHA, slotted-based, spreading-based, and preamble-based methods \cite{Yury2024,Ozates2026CST}, and other promising applications \cite{Zhang2026JSAC,Zhang2025WCL,Gao2026TWC}, where the close-formed achievability bound and converse bound of both conventional antenna and reconfigurable antenna systems can be found in \cite{Zhang2026JSAC} and the URA-enabled integrated sensing and communication system's performance limits can be referred at \cite{Zhang2025WCL}. Generally, slotted-ALOHA refers to the $T$-fold Aloha structure with joint user decoding \cite{Pradhan2019GC,Vem2019TCOM,Marshakov2019VTC,Kowshik2019ISIT,yury2019ACSSC,Kowshik2020TCOM,Andreev2020ISIT}. While slotted-based methods deploy a divide-and-conquer strategy combined with the \ac{CS} algorithm \cite{Liu2018tsp} to reduce the codebook dimension via tree codes \cite{Amal2020TIT} and their variants \cite{SPARC2021TIT,Feng2021TIT,xie2022TCOM,Shyianov2021JSAC,Che2022JSAC}. Due to the limited spectral efficiency of the slotted-based architecture, preamble-based and spreading-based schemes have been extensively investigated and achieved substantial progress. Specifically, spreading-based methods mitigate multi-user interference via random sequence spreading and are typically combined with polar codes \cite{Pradhan2020ICC, Zheng2020VTC, Gkag2023TCOM} or tensor modulation \cite{Decurninge2021WCL, Decu2022GC}. Particularly, the fading-spread \ac{URA} (FASURA) scheme \cite{Gkag2023TCOM} approaches the bound in \cite{Gao2023WCL,Gao2023TIT} and has emerged as a state-of-the-art \ac{MIMO}-\ac{URA} solution. While preamble-based methods partition the data into preamble and payload parts, enabling codeword mapping within a moderate-size codebook to reduce the complexity. Consequently, a two-stage scheme, namely sparse \ac{IDMA}, was proposed in \cite{Pradhan2022TCOM} to effectively mitigate multi-user interference, and has since been further refined in subsequent studies \cite{li2022JSAC, li2025TWC, Ahma2024TWC, Zhang2024TWC, Zhang2024TWC2,Ozates2024TWC}.

\par Notably, sparse \ac{IDMA}, characterized by high spectral efficiency, low complexity, and flexible configurations (e.g., coding schemes and iterative frameworks), has stimulated further research on \ac{URA}. In this context, the \ac{ODMA} architecture, first proposed in \cite{Song2020TCOM}, emerges as a promising direction. It replaces repetition (spreading) and interleaving with idling to generate highly sparse transmission patterns (e.g., time hopping), where idling exhibits superior EXIT performance in multi-user iterative decoding \cite{Song2020TCOM}. Leveraging this advantage, subsequent studies have further investigated the application of \ac{ODMA} framework in \ac{UMA} and \ac{URA} scenarios \cite{Yan2023WCL, Yan2024ISIT, Yan2025TCOM, Ozates2025VTC, Ozates2025SPAWC, Ozates2024WCL, Ozates2024GC, Zhang2024TIOJ, Zhang2025TVT, Zhang2025VTC}, where it consistently achieves notable performance gains compared to its counterpart sparse \ac{IDMA} and other competing frameworks.In general, existing \ac{ODMA} schemes can be broadly classified into three categories: pilot-free, pilot-coupled, and pilot-uncoupled, based on the coding framework and data mapping rules \cite{Zhang2025VTC}. Inherently, the pilot-free scheme performs pattern and data detection without relying on pilots for channel or parameter estimation, thereby achieving relatively high efficiency. This architecture is well-suited for the \ac{GMAC} scenario, as demonstrated in \cite{Yan2023WCL, Yan2024ISIT, Yan2025TCOM}, the proposed \ac{UMA}-\ac{ODMA} framework maps a portion of data into an on-off pattern for user differentiation and also serving as a time-hopping transmission pattern for the remaining data. While the pilot-coupled/uncoupled frameworks are employed in scenarios such as Rayleigh fading or \ac{MIMO} channels where channel estimation is required. In the pilot-coupled scheme specifically, both the pilot and the on-off pattern are derived from the same data segment. Once the active pilots are detected, the transmission patterns can be recovered simultaneously \cite{Ozates2024WCL,Ozates2024GC}, which reduces receiver complexity and improves the reliability of pattern detection. Besides, the joint design for pattern and data detection and channel estimation for \ac{URA} in both \ac{GMAC} and \ac{MIMO} channel scenarios can be found in \cite{Zhang2026TWC}. In contrast, the pilot-uncoupled framework maps the pilot and the on-off pattern from different data segments, such that the detection of active pilots and the recovery of the transmission pattern are performed separately \cite{Zhang2024TIOJ,Zhang2025TVT}. This framework embeds part of the data into the transmission pattern to improve the coding rate; however, it also introduces challenges for pattern detection, which is critical to ensuring accurate data decoding.

\par In the above \ac{ODMA}-related studies, the coding schemes are carefully designed to enhance overall system performance. For example, the authors in \cite{Yan2023WCL,Yan2024ISIT,Yan2025TCOM,Zhang2025VTC} proposed joint data and pattern recovery algorithms for \ac{GMAC}, employing repeat-accumulate or \ac{LDPC} codes to exploit their \ac{SO} information for updating the posterior \ac{LLR} of the transmission pattern. In addition, the incorporation of \ac{LDPC} \ac{SO} information within a \ac{BP}-based iterative framework has also been investigated in other \ac{URA} works \cite{li2022JSAC,Kowshik2019ISIT,yury2019ACSSC,Kowshik2020TCOM}. Moreover, leveraging the advantages of polar codes in the short-blocklength regime, the authors in \cite{Ozates2024WCL, Ozates2024GC} proposed \ac{SIC}-based polar coding schemes for Rayleigh fading and \ac{MIMO} channels, respectively, where an efficient \ac{OMP} algorithm is adopted for channel estimation and pattern detection, yielding significant performance gains over conventional \ac{URA} counterparts. Furthermore, a \ac{TIN}-based polar coding scheme is proposed in \cite{Zhang2024TIOJ} for iterative pattern estimation, while a probabilistic approach is introduced in \cite{Zhang2025TVT} to further reduce the computational complexity. In addition, polar-coded \ac{URA} schemes employing \ac{SC} \cite{SC} or \ac{SCL} \cite{SCL} decoding algorithms have also been extensively studied in \cite{Andreev2020ISIT,Gkag2023TCOM,Ahma2024TWC,Ozates2024TWC,Fengler2022JSAC,Pradhan2020ICC,Zheng2020VTC}.

\par Note that although polar codes outperform \ac{LDPC} codes in the short-blocklength regime and are thus widely adopted, the hard-decision output limits the integration into iterative decoding, thereby constraining pattern detection accuracy and overall system performance. In contrast, despite slightly inferior decoding performance in this regime, \ac{LDPC} codes provide \ac{SO} information that enables iterative processing, effectively enhancing pattern detection accuracy and underpinning the joint pattern and data recovery \cite{Yan2023WCL,Yan2024ISIT,Yan2025TCOM,Zhang2025VTC}. Although \cite{Gkag2023TCOM} proposed a \ac{HO}-based data-aided channel estimation algorithm based on polar codes, it still decouples detection and decoding and remains susceptible to error propagation. To enable the incorporation of polar codes into iterative decoding, prior studies have developed \ac{BP}-based \ac{SO} polar decoding algorithms \cite{BP-Polar-1,BP-Polar-2}. While these approaches provide performance improvements over \ac{SC} decoding, a noticeable gap remains compared to the tailored \ac{HO}-\ac{SCL} polar decoding algorithm. Recently, an \ac{SO}-\ac{SCL} decoding algorithm for polar codes has been preliminarily investigated in \cite{Yuan2025TIT}, which delivers more accurate extrinsic information than the approximation in \cite{Pyndiah1998TWC} and lays the foundation for \ac{SCL}-based bit-wise iterative polar decoding in multi-user transmission systems.

\par Building on the above state-of-the-art review, the \ac{ODMA} framework employs sparse on-off pattern mapping to mitigate user interference, making it a promising transmission architecture for \ac{URA} systems. Accordingly, accurate pattern detection is of critical importance and has become a central focus of existing studies \cite{Yan2023WCL,Ozates2024WCL,Yan2024ISIT,Zhang2024TIOJ,Ozates2024GC,Yan2025TCOM,Zhang2025VTC,Zhang2025TVT}. In pilot-free or pilot-uncoupled scenarios, existing works typically employ iterative pattern detection aided by the \ac{SO} information of repeat-accumulate or \ac{LDPC} decoding to improve detection accuracy \cite{Yan2023WCL,Yan2024ISIT,Yan2025TCOM,Zhang2025VTC}, while the decoding performance remains inferior to that of polar codes in the short-blocklength regime. In pilot-coupled scenarios, existing studies adopt a two-stage framework combining \ac{OMP}-based pattern detection and polar decoding \cite{Ozates2024GC,Ozates2024WCL}, achieving significant gains over other \ac{URA} counterparts. Nevertheless, a noticeable performance gap still exists compared with pilot-uncoupled architectures \cite{Zhang2025TVT}. Overall, in \ac{ODMA}-\ac{URA} scenarios, jointly enhancing pattern detection accuracy and data decoding performance remains an important topic deserving further investigation.

Motivated by the above observations, we consider a pilot-uncoupled \ac{ODMA} framework for \ac{MIMO}-\ac{URA} systems and propose an iterative transmission pattern detection scheme within an \ac{MP} framework assisted by the \ac{SO} information of an \ac{SCL}-based polar decoder. By exploiting the soft information provided by polar codes, the proposed scheme substantially improves pattern detection accuracy. Consequently, the proposed architecture simultaneously achieves both the iterative decoding gain of \ac{SO} polar decoding and detection gain of \ac{SO}-aided pattern detection. The main contributions are summarized as follows:

\begin{itemize}

	\item We investigate a pilot-uncoupled \ac{ODMA} framework for the \ac{MIMO}-\ac{URA} system, where a three-segment transmission architecture is adopted for pilot and interleaving pattern mapping, transmission pattern mapping, and modulated/coded data transmission over on-off patterns, respectively. Compared with the \ac{IDMA} architecture, the adopted framework embeds part of the data into the transmission pattern, thereby reducing the payload size and lowering the coding rate while maintaining the same transmission overhead. Consequently, the effective energy per information bit is increased, leading to improved overall system performance.

	\item We develop an \ac{MP}-based iterative pattern detection algorithm and propose a hierarchical detection strategy to reduce computational complexity while improving detection accuracy. Specifically, \ac{MMSE} estimation and correlation operations are first employed for coarse pattern detection to determine an initial candidate set. Subsequently, the posterior probabilities of transmitted symbols are iteratively computed via \ac{MPA} to update the likelihoods of transmission patterns, based on which a refined candidate set is obtained through \ac{MAP} estimation. Finally, an iterative pattern detection algorithm assisted by the \ac{SO} information of an \ac{SCL}-based polar decoder is proposed to further refine the symbol posterior probability and pattern likelihood, significantly enhancing detection accuracy. The proposed multi-stage strategy progressively narrows the pattern search space, achieving substantial complexity reduction while  substantially improving detection performance.

	\item We establish a \ac{SO} polar-coded \ac{JDD} framework for \ac{MIMO}-\ac{URA} systems. Specifically, the proposed \ac{JDD} framework exploits the bit-wise posterior information of transmitted symbols and iteratively exchanges and updates it between the multi-user detector and the \ac{SO} \ac{SCL}-based polar decoder via \ac{MPA}, overcoming the technical barrier that has long hindered the integration of iterative decoding with polar codes. Essentially, the proposed framework simultaneously leverages the superior error-correction capability of polar codes in the short-blocklength regime and the performance gains enabled by iterative decoding framework, thereby enhancing the overall transmission performance. Simulation results demonstrate that the proposed scheme exhibits superior capability in mitigating multi-user interference and outperforms state-of-the-art schemes in the \ac{PUPE} regime of $10^{-3}$.

\end{itemize}

The remaining of this paper is organized as follows. Section \ref{sec-2} presents the system model. Section \ref{sec-3} introduces the encoding scheme. Section \ref{sec-4} elaborates on the decoding scheme including the pattern detection, data decoding, and the joint pattern detection and polar decoding algorithm. Section \ref{sec-5} presents numerical results and Section \ref{sec-6} concludes the paper.

\textit{Notations:} Throughout this paper, scalars, vectors, and matrices are denoted by lowercase, boldface lowercase, and boldface uppercase, respectively. The transpose, conjugate, and conjugate transpose operations are denoted by $\left( \cdot\right)^T, \left( \cdot\right)^*, \left( \cdot\right)^H$, respectively. $\left\| \mathbf{x} \right\|_p$ and $\left\| \mathbf{A} \right\|_F$ denote the standard $l_p$ and Frobenius norms, respectively. $\mathbf{A}[\mathcal{X},:]$ denotes the submatrix of $\mathbf{A}$ formed by extracting the rows indexed by the elements in the set $\mathcal{X}$. $\text{diag} \left\lbrace \mathbf{d} \right\rbrace$ denotes a diagonal matrix with the vector $\mathbf{d}$ being the diagonals. $\left[T\right]$ denotes the set of integers from $1$ to $T$. $|\mathcal{X}|$ denotes cardinality of  the set $\mathcal{X}$. $\mathcal{CN}(x;\mu,\sigma^2)$ denotes the complex Gaussian distribution of a random variable $x$ with mean $\mu$ and variance $\sigma^2$. $\text{Tr}(\cdot)$ denotes the trace of a matrix. $\mathbb{C}$ and $\mathbb{Z}$ denote the sets of all complex numbers and integers, respectively. $\propto$ denotes the direct proportionality.

\section{System Model} \label{sec-2}

In this paper, we investigate the uplink transmission of a single-cell cellular network consisting of $K_{\rm tot}$ potential single-antenna users, where the signals are assumed to be synchronized in both time and frequency domains. In typical massive random access scenarios, the traffic is sporadic \cite{Liu2018tsp}, meaning that, at any given time, only a small number $K_a$ (with $K_a \ll K_{\rm tot}$) of users are active and transmit data to the \ac{BS}. As commonly assumed in many random access studies \cite{Amal2020TIT,Feng2021TIT,xie2022TCOM,Shyianov2021JSAC,Che2022JSAC,Liu2018tsp,Pradhan2022TCOM,li2022JSAC,li2025TWC,Ahma2024TWC,Zhang2024TWC,Zhang2024TWC2,Ozates2024WCL,Zhang2024TIOJ,Ozates2024GC}, $K_a$ is fixed but unknown to the receiver. The receiver first estimates the number $K_a$ of active users (or equivalently, the number of active codewords) via activity detection, and then performs the subsequent channel estimation \footnote{ In some studies \cite{Liu2018tsp,Zhang2026JSTSP}, activity detection and channel estimation are jointly performed by adopting a Bayesian modeling framework and estimating the equivalent channel based on \textit{maximum a posteriori} estimation, which is commonly referred to as the \ac{JADCE} paradigm.} and data decoding. The \ac{BS} is equipped with $M$ antennas, and each user transmits $B$-bit data over a resource block of length $n$. In this paper, we consider a quasi-static Rayleigh fading channel model, in which the channel coefficients remain constant over the entire duration of a codeword transmission \cite{Gkag2023TCOM,Fengler2022JSAC,Ozates2024GC,Kowshik2019ISIT,yury2019ACSSC,Kowshik2020TCOM,Andreev2020ISIT}. Let $\mathbf{h}_k \in \mathbb{C}^{M\times 1}$ denote the channel vector between user $k$ and the $M$ received antennas, the entries of which are assumed to be \ac{i.i.d.}, following a complex Gaussian distribution with zero mean and unit variance.

Considering the \ac{URA} scenario, let $\mathbf{v}_k \in \{0,1\}^{B\times 1}$ denote the $B$-bit data of user $k$, which is mapped to the transmitted signal $\mathbf{x}\in\mathbb{C}^{n\times 1}$ through a series of user-independent operations, such as encoding, modulation, and codeword mapping. These operations are identical for all users at the transmitters and cannot be distinguished by the \ac{BS} across different users, thus leading to the so-called \ac{URA} nature. Correspondingly, the received signal $\mathbf{Y}\in\mathbf{C}^{n\times M}$ at the \ac{BS} is given by
\begin{equation}
	\mathbf{Y} = \sum\nolimits_{k\in \mathcal{K}_a} \mathbf{x}_k \mathbf{h}_k^T + \mathbf{Z} = \mathbf{X}\mathbf{H}^T + \mathbf{Z} \label{equ-1}
\end{equation}
where the set $\mathcal{K}_a$ includes all active users. Matrices $\mathbf{X} = [\mathbf{x}_1, \mathbf{x}_2, \cdots, \mathbf{x}_{K_a}] \in \mathbb{C}^{n\times K_a}$ and $\mathbf{H} = [\mathbf{h}_1, \mathbf{h}_2, \cdots, \mathbf{h}_{K_a}] \in \mathbb{C}^{M\times K_a}$ denote the transmitted signals and channel coefficients, respectively. And the matrix $\mathbf{Z}\in\mathbb{C}^{n\times M}$ denotes the \ac{AWGN}, where the entries are \ac{i.i.d.} following $\mathcal{CN}(0,\sigma_n^2)$. Assuming the symbols are transmitted with the unit power, the energy-per-bit to the noise power spectral density ratio of the system is defined as
\begin{equation}
	\frac{E_b}{N_0} = \frac{\|\mathbf{x}_k\|_2^2}{B\sigma_n^2} < \frac{1}{R\sigma_n^2} \label{equ-2}
\end{equation}
where $R=B\slash n$ denotes the coding rate, and the inequality is due to the existence of idle transmitted symbols \footnote{ Under the \ac{ODMA} framework, the codeword $\mathbf{x}_k$ contains many ``off'' entries with zero values, while the ``on" channel positions are used to transmit symbols with unit power, as detailed in Eqs. \eqref{equ-6} and \eqref{equ-7}. Therefore, the total energy of the codeword $\mathbf{x}_k$, i.e., $\|\mathbf{x}_k\|_2^2$, is less than $n$.}, as will be discussed in Sec. \ref{sec-3}. In \ac{URA}, the \ac{BS} is tasked to recover a list of transmitted messages $\mathcal{L}(\mathbf{Y})$ based on the received signal $\mathbf{Y}$, where the system performance is evaluated by the probability of missed detection $P_{\rm md}$ and false alarm $P_{\rm fa}$, given by
\begin{align}  
	 P_{\mathrm{md}} &= \frac{1}{{{K_a}}}\sum\nolimits_{k \in {\mathcal{K}_a}} \mathbb{E}[{P\left( {{\mathbf{v}_{{k}}} \notin \mathcal{L}(\mathbf{Y})} \right)}],  \label{equ-3} \\ 
	 P_{\mathrm{fa}} &= \frac{\mathbb{E}[{\left|  {\mathcal{L}(\mathbf{Y})\backslash \left\{ {{\mathbf{v}_{{k}}}:k \in {\mathcal{K}_a}} \right\}} \right|}]}{{\left| \mathcal{L}(\mathbf{Y})\right|}} \label{equ-4}
\end{align} 
and the \ac{PUPE} of the system, $P_e$, is defined as \cite{Gkag2023TCOM}
\begin{equation}
	P_e = P_{\mathrm{md}}+ P_{\mathrm{fa}}. \label{equ-5}
\end{equation}
The objective of this work is to enhance the transmission performance and energy efficiency of the URA system. Specifically, it aims to minimize the \ac{PUPE} $P_e$ for a given ${E_b}\slash{N_0}$, and to minimize the required ${E_b}\slash{N_0}$ for the targeted error rate $\epsilon$ such that $P_e < \epsilon$.

\begin{figure*}[t]
	\centerline{\includegraphics[width=0.9\textwidth]{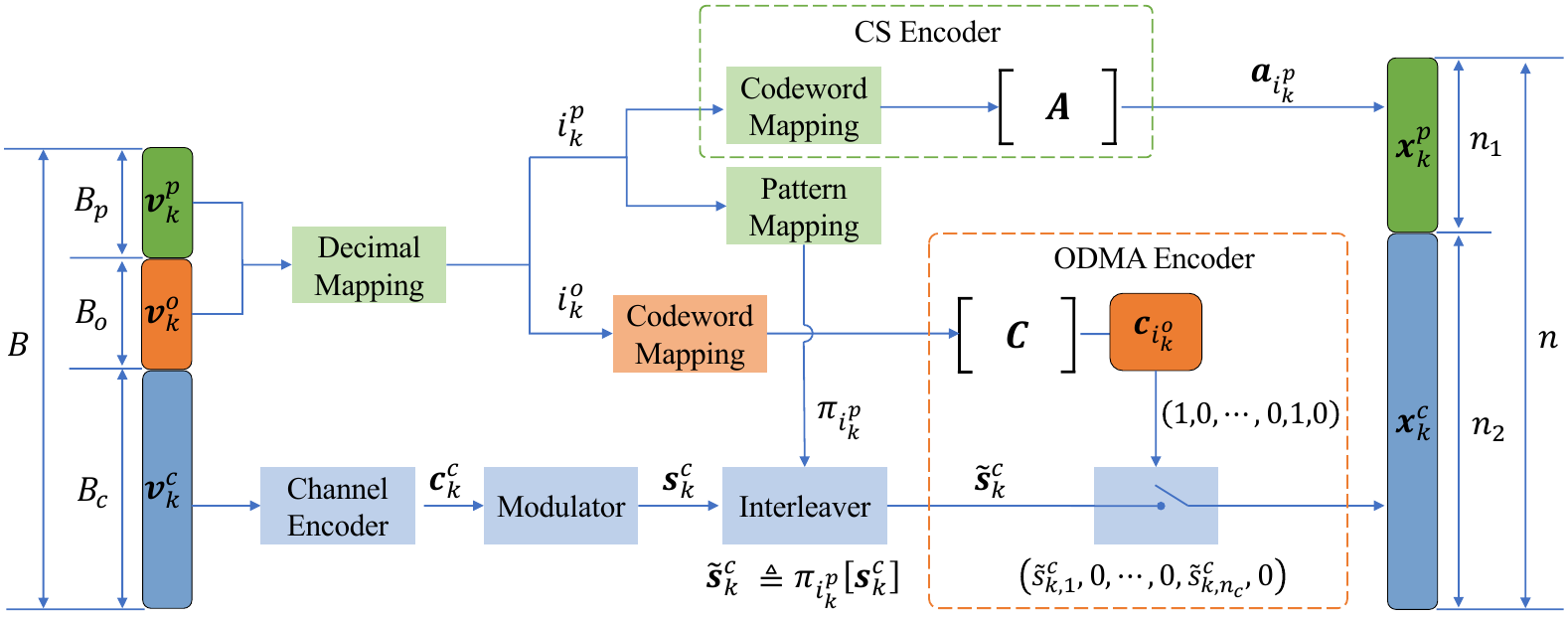}}
	\caption{The overall system encoding scheme within the \ac{ODMA} framework.}
	\label{pic-1}
\end{figure*}

\section{Encoding Scheme} \label{sec-3}

\par In this section, we elaborate on the encoding scheme of the \ac{MIMO}-\ac{URA} system within the \ac{ODMA} framework. Specifically, each user's data is partitioned into three segments, namely $ \mathbf{v}_k^p \in \{0,1\}^{B_p\times1} $, $ \mathbf{v}_k^o \in \{0,1\}^{B_o\times1} $, and $ \mathbf{v}_k^c \in \{0,1\}^{B_c\times1} $, such that a complete bit stream is given by $\mathbf{v}_k = [(\mathbf{v}_k^p)^T, (\mathbf{v}_k^o)^T, (\mathbf{v}_k^c)^T]^T$. These three components are designed to serve distinct functions, as detailed below. 

\par As illustrated in Fig. \ref{pic-1}, the first $B_p$ bits, i.e., $\mathbf{v}_k^p$, function as the preamble, which is primarily used for channel estimation and the transmission of control information (e.g., the interleaving pattern). The encoding of $\mathbf{v}_k^p$ involves mapping it to a decimal index $i_k^p \triangleq \mathrm{dec}[\mathbf{v}_k^p] + 1$, and selecting the $i_k^p$-th column from the Gaussian codebook $\mathbf{A} \in \mathbb{C}^{n_1 \times N_p}$ with \ac{i.i.d.} items, where $N_p = 2^{B_p}$, and $\mathrm{dec}[\cdot]$ denotes the decimal representation of a binary vector. We further defined $\mathbf{c}_k^p \in \{0,1\}^{N_p \times 1}$ as a one-hot binary selection vector with a single $1$ at position $i_k^p$ and zeros elsewhere. Then the resulting preamble sequence is thus $\mathbf{x}_k^p = \mathbf{A}\mathbf{c}_k^p \in \mathbb{C}^{n_1 \times 1}$. 

\par For the second data segment, the following $B_o$-bit of data $\mathbf{v}_k^o$ is employed to select the on-off pattern sequence from the codebook $\mathbf{C}\in \{0,1\}^{n_2 \times N_o}$, where $N_o = 2^{B_o}$ and $n_1+n_2=n$. The codebook $\mathbf{C}$ is constructed as a binary regular sparse matrix, where the Hamming weights of the column and row are $n_c$ and $n_c N_o /n_2$, respectively. Similar to the encoding procedure of $\mathbf{v}_k^p$, the bit sequence $\mathbf{v}_k^o$ is first converted to a decimal index $i_k^o \triangleq \mathrm{dec}[\mathbf{v}_k^o] + 1$, and the corresponding on-off pattern is then given by $\mathbf{x}_k^o = \mathbf{C} \mathbf{c}_k^o \in \{0,1\}^{n_2 \times 1}$, where $\mathbf{c}_k^o \in \{0,1\}^{N_o \times 1}$ is a one-hot vector with its only non-zero entry at index $i_k^o$. Note that although both $\mathbf{v}_k^p$ and $\mathbf{v}_k^o$ are mapped to codewords in the codebook (i.e., $\mathbf{A}$ and $\mathbf{C}$), their roles are fundamentally different. The codeword corresponding to $\mathbf{v}_k^p$ is physically transmitted and serves as a preamble for \ac{JADCE} function. In contrast, the codeword associated with $\mathbf{v}_k^o$ is not transmitted but only specifies the transmission pattern, indicating which $n_c$ out of $n_2$ channel uses are allocated to the third data segment. While $\mathbf{v}_k^o$ can be effectively recovered by the joint pattern detection and data decoding algorithm, as will be elaborate in Sec. \ref{sec-4}. 

\par While for the third data segment, the final $B_c$ bits, i.e., $\mathbf{v}_k^c$, constitute the coding part and carries the primary data payload. The sequence $\mathbf{v}_k^c$ is first encoded using a \ac{CRC} code of length $B_{\mathrm{crc}}$ to enable \ac{SCL} decoding at the receiver. The CRC-augmented bit stream is subsequently encoded using  an $(n_c,B_c+B_{\rm crc})$ polar code, and the resulting codeword is modulated via the BPSK modulation to yield the sequence $\mathbf{s}_k^c \in \mathbb{C}^{n_c \times 1}$. Next, $\mathbf{s}_k^c$ is interleaved to $ \tilde{\mathbf{s}}_k^c = \pi_{i_k^p}[\mathbf{s}_k^c]$ according to the pattern  determined by $\mathbf{v}_k^p$, where $\pi_{i_k^p}[\cdot]$ denotes the interleaver with the pattern of index $i_k^p$. Note that the introduced interleaving operation may appear somewhat inconsistent with the original \ac{ODMA} concept in \cite{Song2020TCOM}, where the idling mechanism replaces interleaving and repetition operations. However, they still fundamentally share the same underlying principle. Specifically, dedicated data-mapping patterns are employed to realize channel idling, whereas the interleaving operation is coupled with the preamble associated with the first data segment and is therefore independent of the \ac{ODMA} architecture. Since the adopted on-off mapping does not alter the relative positions of symbols, the nonlinear characteristics introduced by interleaving can further mitigate inter-user interference during multi-user decoding, thereby improving the decoding performance. Moreover, such an interleaving operation is also commonly employed in multi-user coding schemes. Note that the interleaver and on-off pattern serve different purposes in the proposed coding scheme. Specifically, the on-off pattern introduces sparsity to reduce inter-user interference and improve transmission performance, whereas the interleaver randomly permutes the encoded and modulated data to introduce nonlinear characteristics, thereby enhancing multi-user decoding performance. The interleaved symbols are then mapped to a subset of channel uses specified by the on-off pattern $\mathbf{x}_k^o$ determined by second data segment $\mathbf{v}_k^o$. Define the index set of active channel uses as $\mathcal{N}_c = \{ i\in [n_2] \mid \mathbf{x}_k^o(i) = 1\}$; since $\mathbf{x}_k^o$ has Hamming weight $n_c$, exactly $n_c$ out of $n_2$ entries are non-zero and correspond to the ``on” channel uses allocated for data transmission, while the remaining $n_2 - n_c$ ``off” positions are left idle. Let $\mathbf{x}_k^c \in \mathbb{C}^{n_2 \times 1}$ denote the final coded sequence for the third segment. Then, we have
\begin{equation}
	\mathbf{x}_k^c(i) = \left\lbrace 
	\begin{array}{l} 
		\tilde{\mathbf{s}}_k^c(j), ~~ i\in \mathcal{N}_c, j ~\text{is the index of} ~i~ \text{in} ~ \mathcal{N}_c, \\
		0, ~~ \text{otherwise}.
	\end{array} \right. \label{equ-6}
\end{equation}
Finally, the encoded sequence of user $k$, i.e., $\mathbf{x}_k\in \mathbb{C}^{n\times 1}$ is given by
\begin{equation}
	\mathbf{x}_k = [(\mathbf{x}_k^p)^T, (\mathbf{x}_k^c)^T]^T. \label{equ-7}
\end{equation}
Note that all the aforementioned encoding operations are user-independent and therefore contain no user-specific information. The subscript $k$ in the variables merely indicates the encoding instance for user $k$, and the same procedure is applied to all users in $\mathcal{K}_a$. Consequently, the \ac{BS} only needs to recover the set of transmitted messages without associating them with individual users.

\par The above encoding process is kind of a variant of  existing sparse \ac{IDMA} schemes \cite{Pradhan2022TCOM,li2022JSAC}, which also alleviate multi-user interference through sparse user-specific mappings. However, a key distinction lies in that additional data is explicitly utilized to control the on-off pattern in the adopted \ac{ODMA} framework. Particularly, the data segment $\mathbf{v}_k^o$ is implicitly conveyed through the on-off pattern embedded in the third data segment.  This design effectively reduces the coding rate of the data segment while maintaining the same transmission overhead. As a result, the lower coding rate yields a polar decoding gain, and the increased effective energy per information bit (i.e., $E_b\slash N_0$) further enhances the overall system performance. This performance gain is assessed and demonstrated in the simulation results presented in Sec. \ref{sec-5}.

\begin{Remark}
	 Notably, the scheme employing dedicated data to independently map the transmission pattern is referred to as the pilot-uncoupled \ac{ODMA} architecture, which is first proposed and studied in \cite{Zhang2024TIOJ,Zhang2025TVT}. In this framework, the transmission pattern is decoupled from pilot selection, making the pattern recovery algorithm design a central focus of this study. Although we share a similar uncoupled framework with \cite{Zhang2024TIOJ,Zhang2025TVT}, the proposed pattern detection algorithm fundamentally differs: in the aforementioned works, pattern detection and data decoding are performed separately, without leveraging the posterior information from the decoder. In contrast, as detailed in the following section, the proposed scheme exploits the \ac{SO} information of the \ac{SCL}-based polar decoder and introduces a hierarchical and \ac{MP}-based iterative pattern detection algorithm, effectively reducing detection complexity and substantially improving pattern recovery accuracy.
\end{Remark}

\section{Decoding Scheme} \label{sec-4}

\par In this section, we first introduce the proposed \ac{MMV}-\ac{AMP}-based \ac{JADCE} algorithm for channel estimation and the recovery of the first data segment. Subsequently, we elaborate on the proposed correlation-based pattern detection, \ac{MP}-based pattern detection, and joint pattern detection and data decoding algorithms, which are described in detail in the following subsections, respectively.

\subsection{Joint Activity Detection and Channel Estimation} \label{sec-4-1}

\par As discussed above, the encoding of the first data segment primarily serves to enable channel estimation and the retrieval of control parameters, such as the number of active users and interleaving patterns. Accordingly, the received signal corresponding to this segment is rewritten as follows:

\begin{equation}
		\mathbf{Y}_p= \sum\nolimits_{k\in\mathcal{K}_a} \mathbf{A}\mathbf{c}_k^p\mathbf{h}_k^T + \mathbf{Z} = \mathbf{AC}^p \mathbf{H}^T +\mathbf{Z} \label{equ-8}
\end{equation}
where the matrix $\mathbf{C}^p = [\mathbf{c}_1^p,\cdots,\mathbf{c}_{K_a}^p]\in\{0,1\}^{N_p\times K_a}$ aggregates the selection vectors of all active users. Note that $\mathbf{C}^p$ is an ultra-sparse matrix, where each column contains exactly one non-zero entry. Define the equivalent channel matrix as $\mathbf{H}_e\triangleq \mathbf{C}^p\mathbf{H}^T\in\mathbb{C}^{N_p\times M}$, which exhibits row sparsity due to the condition $N_p \gg K_a$. Correspondingly, the recovery of $\mathbf{H}_e$ can be formulated as a classical \ac{CS} problem. In this context, identifying the support of the non-zero rows of $\mathbf{H}_e$ and estimating the corresponding values constitute the \ac{JADCE} problem \cite{Liu2018tsp}. In this paper, we adopt the \ac{AMP} algorithm to accomplish the aforementioned tasks. In the context of a \ac{MIMO} system, the recovery of $\mathbf{H}_e$ is formulated as a \ac{MMV} problem, where the \ac{MMV}-\ac{AMP} algorithm utilizes a vector denoiser that operates on each row vector of the matched ﬁlter output \cite{Liu2018tsp}. Let $\hat{\mathbf{H}}_e^t = [\hat{\mathbf{h}}_{e,1}^t,\hat{\mathbf{h}}_{e,2}^t,\cdots, \hat{\mathbf{h}}_{e,N_p}^t]^T\in \mathbb{C}^{N_p\times M}$ denote the estimated result of $\mathbf{H}_e$ at the $t$-th iteration. Based on this, the iterative process of the \ac{MMV}-\ac{AMP} algorithm is given by \cite{Liu2018tsp}
\begin{align}
    \hat{\mathbf{h}}_{e,n}^{t+1} &= \eta_{t,n}((\mathbf{R}^t)^H\mathbf{a}_n + \hat{\mathbf{h}}_{e,n}^t),\label{equ-9} \\
    \mathbf{R}^{t+1} &= \mathbf{Y}_p - \mathbf{A}\hat{\mathbf{H}}_e^{t+1} + \frac{N_p}{n_1}\mathbf{R}^t \sum_{n=1}^{N_p}{\frac{\eta_{t,n}'((\mathbf{R}^t)^H\mathbf{a}_n + \hat{\mathbf{h}}_{e,n}^t)}{N_p}} \label{equ-10}
\end{align}
where $\mathbf{a}_n$ denotes the $n$-th column of the codebook $\mathbf{A}$, and $\|\mathbf{a}_n\|_2^2 = 1$. The function $\eta_{t,n}(\cdot): \mathbf{C}^{M\times 1} \rightarrow \mathbf{C}^{M\times 1}$ is the vector-wise denoiser and $\eta_{t,n}'(\cdot)$ represents its first order derivative. A \ac{MMSE} denoiser function is employed in \cite{Liu2018tsp}, and the theoretical analysis demonstrates that the probabilities of missed detection and false alarm for the codewords approach zero in the asymptotic regime. Additionally, the channel estimation error will ultimately converge. The iterative process in Eqs. \eqref{equ-9} and \eqref{equ-10} terminates when the maximum number of iterations or a specific criterion is reached, yielding the estimated equivalent channel $\hat{\mathbf{H}}_e$. Correspondingly, define $\mathcal{X} = \{\hat{i}_p^k \mid k \in \hat{\mathcal{K}}_a\}$ as the indices of the non-zero rows of $\hat{\mathbf{H}}_e$, which can be readily identified via energy detection. Here,  $\hat{\mathcal{K}}_a$ denotes the set of estimated active users and $\hat{K}_a = |\hat{\mathcal{K}}_a|$ represents its cardinality. In this context,  the first data segment can be efficiently recovered via reverse mapping, and the corresponding channel estimates are given by $\hat{\mathbf{H}} \triangleq \hat{\mathbf{H}}_e[\mathcal{X},:] \in \mathbb{C}^{\hat{K}_a \times M}$. After obtaining the above preliminary estimates, we will focus on presenting the pattern detection and data decoding schemes in the following contents.

\subsection{Coarse Detection: Correlation-based Pattern Detection} \label{sec-4-2}

\begin{figure}[htpb]
	\centerline{\includegraphics[width=0.35\textwidth]{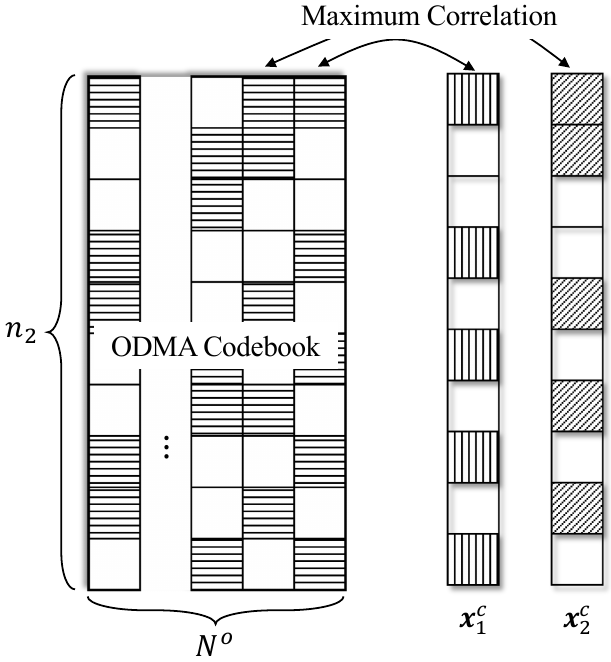}}
	\caption{Identify the matching on-off pattern via correlation operation.}
	\label{pic-2}
\end{figure}

After detecting the active users and estimating the corresponding channels, a key challenge lies in identifying the embedded on-off patterns within the data sequences, i.e., $\mathbf{x}_k^c$. Accurate detection of these patterns is essential for determining the specific channel uses and enabling reliable data decoding. To begin with, the received signal corresponding to the third data segment is given by
\begin{equation}
	\mathbf{Y}_c = \sum\nolimits_{k\in\mathcal{K}_a} \mathbf{x}_k^c \mathbf{h}_k^T + \mathbf{Z}. \label{equ-11}
\end{equation}
In order to detect the active patterns within $\mathbf{x}_k^c$, \cite{Yan2023WCL,Zhang2025VTC} proposed an \ac{MP}-based joint pattern and data recovery algorithm under \ac{GMAC}. This approach iteratively computes the conditional probability \ac{LLR}s for each of the $N_o$ candidate patterns during the \ac{MP} iterations, while the computational complexity remains manageable under the \ac{GMAC} scenario, since the patterns are decoupled from the channel, thereby avoiding additional combinatorial expansions.  However, in the \ac{MIMO} channel scenario considered in this work, the transmission patterns are coupled with the channel, causing the pattern detection search space to expand to $N_o \hat{K}_a$. If joint detection over the transmission patterns of all users is performed, the computational complexity will increase exponentially. To address this issue, the proposed hierarchical pattern detection algorithm first exploits correlation-based operations to preliminarily identify a candidate set of highly correlated transmission patterns, thereby obtaining a coarse-grained pattern detection result and effectively reducing the subsequent search space. Prior to the correlation operation, a \ac{LMMSE} estimator is employed to obtain an estimate of the data sequence, i.e., $\hat{\mathbf{X}}^c = [\hat{\mathbf{x}}^c_1,\hat{\mathbf{x}}^c_2,\cdots, \hat{\mathbf{x}}^c_{\hat{K}_a}]^T\in \mathbb{C}^{\hat{K}_a \times n_2}$, which is given by
\begin{equation}
    \hat{\mathbf{X}}^c = \widehat{\mathbf{H}}^* ( \widehat{\mathbf{H}}^T \widehat{\mathbf{H}}^* + \sigma_n^2 \mathbf{I}_M )^{-1} \mathbf{Y}_c^T.  \label{equ-12}
\end{equation}
 Then, by performing correlation operations between the transmission patterns in the shared ODMA codebook $\mathbf{C}$, and each estimated data sequence $\{\hat{\mathbf{x}}^c_k, k\in [\hat{K}_a]\}$, the most compatible transmission pattern associated with each data sequence can be determined, as illustrated in Fig. \ref{pic-2}.
Therefore, the detected on-off pattern for user $k$ is given by
\begin{equation}
    \hat{i}_k^o = \arg \max_i \mathbf{c}_i^T  \hat{\mathbf{l}}_k^c, \quad i\in[N_o], k\in [\hat{K}_a] \label{equ-13} 
\end{equation}
where $\mathbf{c}_i$ is the $i$-th pattern of the codebook $\mathbf{C}$, and 
\begin{equation}
    \hat{\mathbf{l}}_k^c = \frac{2 {\rm Re}\{\hat{\mathbf{x}}^c_k\}}{\sigma_n^2}, k\in[\hat{K}_a] \label{equ-14}
\end{equation}
denotes the \ac{LLR} of $\hat{\mathbf{x}}^c_k$. To reduce the missed detection probability, we construct a coarse candidate set for each user comprising the top-$N_c$ patterns with the highest correlations, as determined by Eq. \eqref{equ-13}, i.e., $\{\hat{i}_{k,j}^o, k\in [\hat{K}_a], j\in [N_c]\}$. Subsequently, the \ac{MP}-based pattern detection algorithm operates on this coarse-grained candidate set of transmission patterns to compute the posterior \ac{LLR} of each candidate pattern. Based on the resulting posterior probabilities, the transmission pattern corresponding to each data sequence is further determined via \ac{MAP} estimation.

\subsection{Precise Detection: MP-based Pattern Detection} \label{sec-4-3}

\par Building upon the preliminary candidate pattern set obtained via correlation operations, this subsection further employs \ac{MPA} to perform \ac{MAP} estimation over the candidate pattern set, thereby achieving more accurate pattern detection results. Note that in the subsequent detection process, the search space of the patterns is reduced from $N_o\hat{K}_a$ to $N_c \hat{K}_a$.  For instance, in the considered simulation setup, $N_o = 2^{16}$ whereas only $N_c=10$ candidate patterns are retained. In conjunction with the adopted joint detection framework, this hierarchical detection strategy yields a substantial reduction in computational complexity. Specifically, the $(l,m)$-th element of $\mathbf{Y}_c$, i.e., $y_{l,m}$ can be written as 
 \begin{equation}
    \begin{aligned}
        y_{l,m} &= \sum_{k=1}^{{K}_a}{h_{k,m}x^c_{k,l}} + z_{l,m} \\
                &= h_{k,m}x^c_{k,l} + \underbrace{\sum_{k'\in \mathcal{K}_a\backslash k}{h_{k',m}x^c_{k',l}} + z_{l,m}}_{n_{k,l,m}}
    \end{aligned} \label{equ-15}
  \end{equation}
  where the transmitted symbol $x^c_{k,l}$ is valued in $\{0,1,-1\}$ for BPSK modulation. According to the Central Limit Theorem, the variable $n_{k,l,m}$ can be approximated as Gaussian distributed, with its mean $\mathbb{E}[n_{k,l,m}]$ and variance $\text{Var}[n_{k,l,m}]$ given by:
  \begin{align}
        \mathbb{E}[n_{k,l,m}] &= \sum_{k'\in \mathcal{K}_a\backslash k , l'\in \mathcal{L}_{k'}}{h_{k',m}\mathbb{E}[x^c_{k',l'}]} \notag \\
        &= \sum_{k'\in \mathcal{K}_a\backslash k , l'\in \mathcal{L}_{k'}}{h_{k',m}\alpha}, \label{equ-16}  \\       
        \text{Var}[n_{k,l,m}] &= \sum_{k'\in \mathcal{K}_a\backslash k , l'\in \mathcal{L}_{k'}}{|h_{k',m}|^2 \text{Var}[x^c_{k',l'}]} + \sigma_n^2 \notag  \\
        &= \sum_{k'\in \mathcal{K}_a\backslash k , l'\in \mathcal{L}_{k'}}{|h_{k',m}|^2 (1-\alpha^2)} + \sigma_n^2 \label{equ-17}      
\end{align}
where $\alpha\triangleq \tanh(L_{k',l'}^a\slash 2)$, and $\mathcal{L}_{k'} = \{l|c_{i_{k'}^o,l}= 1, l\in[n_2]\}$ denotes the indices of non-idling channel uses in pattern $\mathbf{c}_{i_{k'}^o}$ for user $k'\in \mathcal{K}_a$. And we denote the  posterior \ac{LLR} of symbol $x^c_{k',l'}$ as
\begin{equation}
    L_{k',l'}^a = \log \frac{P(x^c_{k',l'}=1)}{P(x^c_{k',l'}=-1)} \label{equ-18}
\end{equation}
which is initialized to zero. The probability of $y_{l,m}$ conditioned on the transmission pattern and data symbol is given by
\begin{equation}
    \begin{aligned}
        &P(y_{l,m} | h_{k,m}, b_{k,i}=p, x^c_{k,l} = q) \\
        &= \frac{1}{\pi \text{Var}[n_{k,l,m}]} \exp \biggl\{ - \frac{|y_{l,m} - (pq h_{k,m} + \mathbb{E}[n_{k,l,m}]) |^2}{\text{Var}[n_{k,l,m}]}\biggr\}
    \end{aligned} \label{equ-19}
\end{equation}
\begin{table*}[b]
	\begin{equation}
		\begin{aligned}
			L(b_{k,i}) &= \log \frac{P(\mathbf{Y}_c | \mathbf{H}, b_{k,i} =1)}{P(\mathbf{Y}_c | \mathbf{H}, b_{k,i} =0)} = \sum_{l=1}^{L_c}\sum_{m=1}^{M} \log \frac{\sum\limits_{q\in \{0,1,-1\}}P(x^c_{k,l}=q)P(y_{l,m} |  h_{k,m},b_{k,i}=1 , x^c_{k,l}=q)}{P(y_{l,m}| h_{k,m}, b_{k,i}=0,x^c_{k,l}=0)} \\
				&= \sum_{l\in \mathcal{L}_k} \sum_{m=1}^{M} \log \frac{\sum\limits_{q\in \{1,-1\}}P(x^c_{k,l}=q)P(y_{l,m} |  h_{k,m},b_{k,i}=1 , x^c_{k,l}=q)}{P(y_{l,m}| h_{k,m}, b_{k,i}=0,x^c_{k,l}=0)} \\
				&= \sum_{l\in \mathcal{L}_k} \sum_{m=1}^{M} \log \Biggl\{ p_1 \cdot \exp \biggl\{\frac{\Re\{h_{k,m}^{*}(y_{l,m} - E)\}}{V}\biggr\} + \bar{p}_1 \cdot \exp \biggl\{-\frac{\Re\{h_{k,m}^{*}(y_{l,m} - E)\}}{V}\biggr\}  \Biggr\} - \frac{|h_{k,m}|^2}{2V}\\
				&\overset{(a)}{\approx} \sum_{l\in \mathcal{L}_k} \sum_{m=1}^{M} \frac{2\Re\{|h_{k,m}^{*}(y_{l,m} - E)|\} - |h_{k,m}|^2}{2V}.
		\end{aligned} \label{equ-20}
	\end{equation}
\end{table*}
where $i\in [2^{B_o}]$, $p\in \{0,1\}$, $q\in \{0,1,-1\}$. The variable $b_{k,i}$ is a binary indicator, where $b_{k,i} = 1$ if user $k$ selects the $i$-th on-off pattern, and $b_{k,i} = 0$ otherwise. Furthermore, within the candidate set of patterns,  the conditional likelihood ratio that user $k \in \mathcal{K}_a$ selects the $i$-th pattern (where $i \in \{\hat{i}_{k,j}^o,\ j \in [N_c]\}$) is given in Eq. \eqref{equ-20} at the bottom of this page, where $\mathcal{L}_k = \left\{l|c_{i,l}= 1, i \in \{\hat{i}_{k,j}^o, j\in [N_c]\}, l\in[n_2]\right\}$, $p_1 \triangleq P(x^c_{k,l}=1)$, $E \triangleq \mathbb{E}[n_{k,l,m}]$, and $V \triangleq \text{Var}[n_{k,l,m}]$. And the approximation term (a) is obtained based on the relation $\log(\exp^x+\exp^y) \approx \max(x,y)$, where the prior probability $P(x^c_{k,l})$ can be regarded as uniformly distributed during the iterations with negligible impact on the results, thereby simplifying the computation of the conditional probability \ac{LLR} of the pattern \cite{Yan2023WCL}. Note that likelihood-ratio metrics similar to that in Eq. \eqref{equ-20} have also been investigated in \cite{Yan2023WCL,Zhang2025VTC} for pattern detection in \ac{GMAC} scenarios. Different from these works, this paper considers the \ac{MIMO} channel scenario and reformulates the conditional probability computation for transmission patterns accordingly. Moreover, to address the exponentially increasing complexity caused by the coupling between transmission patterns and channel coefficients in \ac{MIMO} systems, a hierarchical pattern detection algorithm is proposed for complexity reduction. In addition, a joint pattern detection and data decoding scheme based on the \ac{SO} information of the \ac{SCL} polar decoder is developed, where the posterior probabilities of transmitted symbols are exploited to iteratively refine the likelihood-ratio metrics of transmission patterns, significantly improving the pattern detection accuracy, as demonstrated in the following subsection.

\par Accordingly, applying the candidate patterns, the extrinsic information $L^e_{k,l}$ passed from $y_{l,m}$ to  $x^c_{k,l}$ is given by:
\begin{equation}
    \begin{aligned}
        L^e_{k,l} &= \sum\limits_{m=1}^{M} \log \frac{P(y_{l,m}| h_{k,m}, b_{k,i}=1, x^c_{k,l}=1)}{P(y_{l,m}| h_{k,m}, b_{k,i}=1, x^c_{k,l}=-1)}\\
                  &= \sum\limits_{m=1}^{M} \frac{2\Re\{h_{k,m}^{*}(y_{l,m} -\mathbb{E}[n_{k,l,m}])\}}{\text{Var}[n_{k,l,m}]}, l\in \mathcal{L}_k
    \end{aligned} \label{equ-21}
\end{equation}
where $\mathbb{E}[n_{k,l,m}]$ and $\text{Var}[n_{k,l,m}]$ can be obtained through Eqs. \eqref{equ-16} and \eqref{equ-17}. Then, the a posteriori probability of the symbol $x^c_{k,l}$ is updated based on the extrinsic information $L^e_{k,l}$ obtained in the observations $y_{l,m}$, which is given by
\begin{equation}
	P(x^c_{k,l}= 1) = \frac{\exp( L^e_{k,l})}{1+\exp( L^e_{k,l})}. \label{equ-22}
\end{equation}

\par The above procedure is iteratively applied to each candidate transmission pattern in the set $\{\hat{i}_{k,j}^o,\ k \in [\hat{K}_a],\ j \in [N_c]\}$. Upon convergence, the \ac{LLR} $L(b_{k,i})$ corresponding to the conditional probability of each pattern is obtained. Note that to reduce the complexity, when computing the pattern probability for the specific user $k \in [\hat{K}_a]$, the patterns of all other users $\hat{\mathcal{K}}_a \setminus {k}$ are fixed to the one with the highest correlation instead of joint update. This procedure is alternated across each user during the iteration. After the \ac{LLR}s of all candidate patterns have been computed, the final estimation results are given by
\begin{equation}
    \hat{i}_k^o = \arg \max_i L(b_{k,i}), i \in \{\hat{i}_{k,j}^o, j\in [N_c]\}, k\in \hat{\mathcal{K}}_a. \label{equ-23}
\end{equation} 
In summary, the pattern detection procedure introduced in this subsection is referred to as the \textit{\ac{MP}-based Pattern Detection Algorithm}, which is formally summarized in Algorithm \ref{alg-1}.

 \begin{algorithm} [htpb]
	\setstretch{1.05}
	\caption{The MP-based Pattern Detection Algorithm}
	\label{alg-1}  
	\begin{algorithmic}[1]	
		\STATE {{\bf Input}: $\mathbf{Y}_c$, $\mathbf{C}$, $\hat{\mathbf{H}}$, $\sigma_n^2$, the candidate pattern set $\hat{\mathcal{I}}_k^o = \{\hat{i}_{k,j}^o, j\in [N_c]\}, k\in \hat{\mathcal{K}}_a$.}\\
		\STATE {{\bf Output}: The pattern detection results $\{\hat{i}_k^o, k\in \hat{\mathcal{K}}_a\}$.}\\
		\STATE{{\bf Initial}: The posterior probability $P(x^c_{k,l}=1)=0.5$.}\\

		\FOR{$k=1$ to $\hat{K}_a$ }
			\FOR{$j=1$ to $N_c$}
				\STATE{$\hat{i}_{k}^o=\hat{\mathcal{I}}_k^o(j)$, $\hat{i}_{k'}^o = \hat{\mathcal{I}}_{k'}^o(1)$, $k'\in \hat{\mathcal{K}}_a \backslash k$, $p\triangleq\hat{i}_{k}^o$.}\\
				\FOR {$\text{iter}=1$ to $t_{{\rm max}}$}
				\STATE{Calculate $P(y_{l,m} | h_{k,m}, b_{k,p}, x_{k,l}^c)$ via Eqs. \eqref{equ-15}-\eqref{equ-19}.}\\
				\STATE{Calculate pattern \ac{LLR} $L(b_{k,p})$ via Eq. \eqref{equ-20}.}\\
				\STATE{Calculate extrinsic information $L_{k,l}^e$ via Eq. \eqref{equ-21}.}\\
				\STATE{Update posterior probabilitiy $P(x^c_{k,l})$ via Eq. \eqref{equ-22}.}\\
				\ENDFOR
			\ENDFOR
			\STATE{MAP estimation: $ \hat{i}_k^o = \arg \max_{p} L(b_{k,p}), \quad p \in \hat{\mathcal{I}}_k^o$.}\\
		\ENDFOR
	\end{algorithmic}  
\end{algorithm}

It is noted that a similar \ac{MAP} estimation of transmission patterns based on Eq. \eqref{equ-23} is also employed in \cite{Zhang2024TIOJ,Zhang2025TVT}. Nevertheless, the \ac{MAP} estimation in those works fundamentally differs from the approach proposed herein. Specifically, in \cite{Zhang2024TIOJ,Zhang2025TVT}, the \ac{MAP} estimation of transmission patterns is performed independently without leveraging the posterior information provided by the decoder. In contrast, the \ac{MAP} estimation in this work is primarily used to prune the pattern search space, thereby reducing computational complexity. Subsequently, in the joint pattern detection and decoding algorithm detailed in the next subsection, the posterior probabilities of the transmission patterns are further refined by incorporating the symbol-wise posterior information produced by the \ac{SCL} polar decoder, leading to improved pattern detection accuracy.

\subsection{Enhanced Detection: Joint Pattern Detection and Data Decoding} \label{sec-4-4}

\par Within the proposed hierarchical pattern detection framework, the precise pattern detection algorithm presented in Alg. \ref{alg-1} is designed to further reduce the candidate pattern set for each user, while the final deterministic pattern detection is accomplished by the joint pattern detection and data decoding algorithm proposed in this subsection. The motivation behind this framework arises from the intrinsic coupling between transmission patterns and channel coefficients in \ac{MIMO} channels. Specifically, unlike the approaches in \cite{Yan2023WCL,Zhang2025VTC} for the \ac{GMAC} scenario, directly performing joint pattern detection and data decoding over all candidate patterns would incur substantial decoding complexity over a large search space, thereby significantly increasing the receiver computational burden. Under the proposed hierarchical framework, however, pattern detection is carried out over a progressively reduced search space, effectively lowering the overall computational complexity.

\par Note that existing \ac{URA} schemes incorporating polar codes predominantly adopt hard-decision decoding \cite{Pradhan2020ICC,Zheng2020VTC,Gkag2023TCOM,Ozates2024WCL,Zhang2024TIOJ,Ozates2024GC}, which limits system performance due to the absence of soft-information exploitation in iterative detection and decoding. To overcome this limitation, an \ac{SO}  \ac{SCL} polar decoding algorithm \cite{Yuan2025TIT} is first introduced in this subsection. Based on the posterior probability information provided by the soft outputs of the polar decoder, an \ac{MP}-based joint pattern detection and data decoding algorithm is further proposed to effectively enhance both pattern detection accuracy and data decoding performance.

As preliminary definitions, we introduce the following symbols and variables. Denote $u^N = [u_1, \cdots, u_N]\in \{0,1\}^{N\times 1}$ as the decoding path (i.e., the codeword) subject to the frozen-bit constraints, and $c^{N} = [c_1,\cdots, c_N] \in \{0,1\}^{N\times 1}$ as the corresponding codeword obtained after polar transform, where $N$ denotes the length of the polar code. Furthermore, let $Q_{C^{N} | Y^{N}}(c^{N} | y^{N})$ denote the auxiliary conditional probability of the codeword $c^N$ given the received signal $y^N$, which can be expressed as \cite{Yuan2025TIT}
\begin{equation}
    Q_{C^{N} | Y^{N}}(c^{N} | y^{N}) \triangleq \prod_{i=1}^{N}P_{C|Y}(c_i|y_i)  \label{equ-24}
\end{equation}
where $P_{C|Y}(c_i|y_i)$ denotes the \ac{APP} of the $i$-th symbol $c_i$ given the received symbol $y_i$. Correspondingly, the codebook probability is defined as $Q_C(y^N) \triangleq \sum_{c^N\in \mathcal{C}} Q_{C^{N} | Y^{N}}(c^{N} | y^{N})$, where $\mathcal{C}$ denotes the codebook of the polar codewords. Similarly, it follows that $ Q_U(y^N)  = \sum_{u^N\in\mathcal{U}} Q_{U^{N} | Y^{N}}(u^{N} | y^{N})$ since $u^N$ and $c^N$ are linked via a bijective mapping in polar transform, where $\mathcal{U}$ denotes the set containing all $u^N$. The computation of $Q_U(y^N)$ involves an exhaustive enumeration of all valid codewords or decoding paths, resulting in prohibitive computational complexity. To this end, an approximate method for computing $Q_U(y^N)$ is given by \cite{Yuan2025TIT}
\begin{equation}
	\begin{aligned}
		Q_U(y^N) &\approx Q_U^*(y^N) \triangleq \sum_{u^N\in\mathcal{V}} Q_{U^{N} | Y^{N}}(u^{N} | y^{N})\\
		 &+ \sum\limits_{w^i \in \mathcal{W}} 2^{-|\mathcal{F}(i:N)|} Q_{U^{i} | Y^{N}}(w^{i} | y^{N})
	\end{aligned} \label{equ-25}
\end{equation}
where $\mathcal{V}$ and $\mathcal{W}$ denote the set of visited leaves and the set of roots of unvisited trees in the \ac{SCL} decoding tree, respectively, and $|\mathcal{F}(i:N)|$ represents the number of frozen bits in the subsequently decoding process. Based on this, the bit-wise \ac{APP} from the \ac{SO} polar decoder is provided in Eq. \eqref{equ-26} at the top of next page \cite{Yuan2025TIT}.

\begin{table*}[t]
	\begin{equation}
			L^a_i \approx  \log \frac{\sum\limits_{c_i=0,c^N \in \mathcal{L}_C} Q_{C^{N} | Y^{N}}(c^{N} | y^{N}) + \big(Q_U^*(y^N) - \sum_{c^N \in \mathcal{L}_C} Q_{C^{N} | Y^{N}}(c^{N} | y^{N})\big) \cdot P_{C|Y}(0|y_i)}{\sum\limits_{c_i=1,c^N \in \mathcal{L}_C} Q_{C^{N} | Y^{N}}(c^{N} | y^{N}) + \big(Q_U^*(y^N) - \sum_{c^N \in \mathcal{L}_C} Q_{C^{N} | Y^{N}}(c^{N} | y^{N})\big) \cdot P_{C|Y}(1|y_i)}. \label{equ-26}
	\end{equation}
\end{table*}

\par Once the \ac{SO} posterior information from the polar decoder is obtained, the iterative data decoding process can be performed based on the \ac{MPA}. At this stage, the transmission pattern to each codeword is fixed according to the \ac{MAP} estimation result in Alg. \ref{alg-1} during the iterative decoding process. In the discussion of \ac{MPA}, we define the received signal $y_{l,m} (l\in[n_2], m\in[M])$ as the \ac{SN}, and the interleaved symbol $\tilde{s}_{k,n} (k\in [K_a], n\in[n_c])$ as the \ac{VN}. Furthermore, the message from \ac{VN} $\tilde{s}_{k,n}$ to \ac{SN} $y_{l,m}$ is denoted by $L^a_{k,n \rightarrow l,m}$, representing the \ac{LLR} of the posterior probability output from the polar decoder, while the message from \ac{SN} $y_{l,m}$ to \ac{VN} $\tilde{s}_{k,n}$ is denoted by $L^e_{l,m \rightarrow k,n}$, representing the \ac{LLR} of the posterior probability observed at $y_{l,m}$. For convenience, we define the set $\mathcal{D}_l$ containing all \ac{VN}s connected to the \ac{SN} $y_{l,m}$ based on the deterministic transmission pattern. The update rule for $L^e$ is then given by
\begin{equation}
	\begin{aligned}
		L^e_{l,m \rightarrow k,n} 
		&= \log \frac{P(y_{l,m} | \mathbf{H},\tilde{s}_{k,n}=1)}{P(y_{l,m} | \mathbf{H},\tilde{s}_{k,n}=-1)} \\
		&= \frac{2}{V}\Re (h_{m,k}^*(y_{l,m}-E))
	\end{aligned} \label{equ-27}
\end{equation}
where $y_{l,m}$ denotes the $(l,m)$-th entry in $\mathbf{Y}_c$, and 
\begin{align}
	E &= \sum\limits_{(i,j)\in \mathcal{D}_l /(k,n)} h_{m,i}  \tanh(\tilde{L}^a_{i,j\rightarrow l,m}/2), \label{equ-28} \\
	V &= \sum\limits_{(i,j)\in \mathcal{D}_l /(k,n)} |h_{m,i}|^2 (1- \tanh^2(\tilde{L}^a_{i,j\rightarrow l,m}/2)) + \sigma_n^2. \label{equ-29}
\end{align}
Note that in the above computation, $\tilde{L}^a$ is obtained by interleaving $L^a$ in Eq. \eqref{equ-26}, with the interleaving pattern determined by the first data segment, which is recovered according to the \ac{MMV}-\ac{AMP} algorithm. Accordingly, the update rule for $L^a$ can be derived from Eq. \eqref{equ-26}. Specifically, for the codeword of user $k$ $\mathbf{c}_k$, we have 
\begin{equation}
	P_{Y|C}(y_{l,m}|c_{k,n}=0) = \frac{\exp(L^e_{l,m \rightarrow k,n})}{1+\exp(L^e_{l,m \rightarrow k,n})} \label{equ-30}
\end{equation}
and given that the transmitted symbols are uniformly distributed, the posterior probability of the codeword $P_{C|Y}(c_{k,n}|y_{l,m})$ can be readily obtained using the Bayes's rule, i.e., 
\begin{equation}
    P_{C|Y}(c_{k,n}|y_{l,m}) = \frac{P_{Y|C}(y_{l,m}|c_{k,n}) \cdot P_C(c_{k,n})}{\sum_{q\in \{0,1\}}P_{Y|C}(y_{l,m}|q)\cdot P_C(q)}. \label{equ-31}
\end{equation}
Correspondingly, based on Eqs. \eqref{equ-24}-\eqref{equ-26} and the \ac{SCL} polar decoding process, the \ac{SO} posterior information of the codeword $s_{k,n}$, i.e., $L^a_{k,n \rightarrow l,m}$ can be obtained. During the iterative process, the posterior \ac{LLR}s $L^e$ and $L^a$ are iteratively updated., terminating when either the maximum iteration count is reached or the decoding result satisfies the \ac{CRC} check. Compared to the conventional \ac{HO} \ac{SCL} polar decoding with hard decision output of the codewords, the proposed algorithm utilizes the decoder's \ac{SO} information and employs an iterative decoding process based on the \ac{MPA}, achieving a substantial improvement in decoding performance, and this performance gain will be  assessed in Sec. \ref{sec-5}.

\par To this point, we have presented the iterative data decoding process for polar codes. Notably, data decoding is conducted based on the pattern detection results in Alg. \ref{alg-1}, implying that errors in pattern detection may propagate to the decoding stage, potentially resulting in decoding failure. Additionally, in Sec. \ref{sec-4-3}, the computation of the pattern's \ac{LLR}s relies solely on the channel extrinsic information, without incorporating the posterior probability from the decoder, thereby limiting the accuracy of pattern detection. To address these challenges, under the proposed hierarchical pattern detection framework, an enhanced pattern detection scheme, namely, the joint pattern detection and data decoding algorithm, is further developed. Building upon Alg. \ref{alg-1}, the proposed algorithm exploits the posterior probability information provided by the soft outputs of the polar decoder to refine and update the posterior probabilities of the transmission patterns, thereby improving the pattern detection accuracy. Meanwhile, the updated pattern detection results are fed back to the decoder for subsequent decoding iterations, which further enhances the data decoding performance.

\begin{figure}[htpb]
	\centerline{\includegraphics[width=0.45\textwidth]{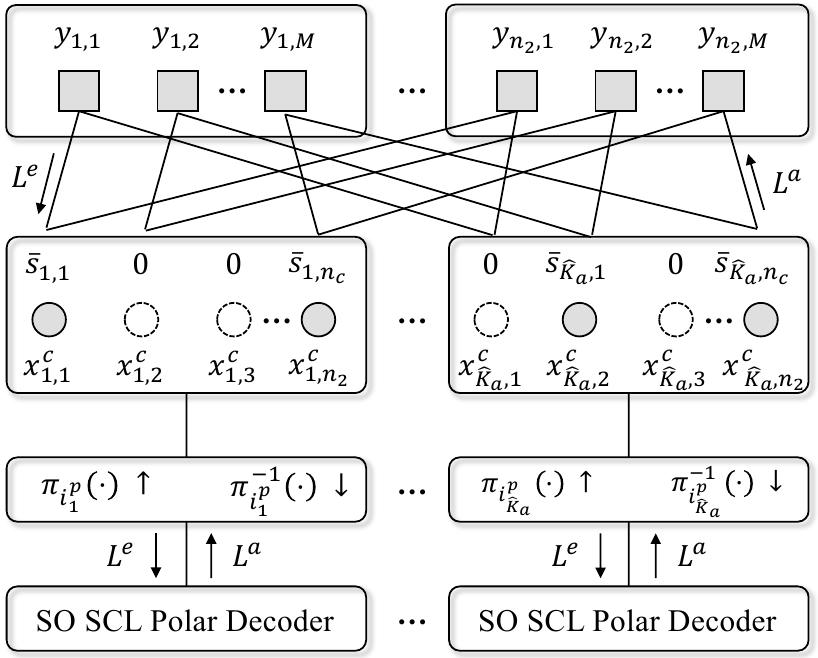}}
	\caption{Schematic diagram of the joint pattern detection and data decoding algorithm.}
	\label{pic-3}
\end{figure}

\par Considering that the posterior \ac{LLR} of the pattern computed from Eq. \eqref{equ-20} does not incorporate the decoder's information, the resulting inaccuracies may cause the pattern with the maximum \ac{LLR} to differ from the actual one selected by the user. To this end, let $\hat{\mathcal{I}}_k^o$ denote the set of the top-$N_d$ patterns with the highest \ac{LLR} values $L(b_{k,i})$ for user $k\in \hat{\mathcal{K}}_a$,  computed in Eq. \eqref{equ-20}, i.e., $\hat{\mathcal{I}}_k^o = \{\hat{i}_{k,j}^o, j\in [N_d]\}$. Correspondingly, for each pattern in $\hat{\mathcal{I}}_k^o$, the iterative decoding process for \ac{SO} polar codes is conducted by exchanging the posterior information given in Eqs. \eqref{equ-26}-\eqref{equ-27}.

\par In the decoding process of a specific user $k$, the proposed joint pattern detection and decoding algorithm is executed $N_d$ times to successively update the \ac{LLR}s of all patterns in the set $\hat{\mathcal{I}}_k^o$. In this process, the pattern of each of the other users, $u \in \hat{\mathcal{K}}_a \backslash k$, is fixed to the first element of its corresponding set $\hat{\mathcal{I}}_u^o$, i.e., the pattern with the highest \ac{LLR}. The overall procedure terminates once the \ac{LLR}s of all patterns across all users, i.e., $\{\hat{\mathcal{I}}_k^o, k\in \hat{\mathcal{K}}_a\}$ have been updated.

\par Subsequently, based on the specific transmission pattern, the decoder outputs the posterior probability \ac{LLR} values according to Eqs. \eqref{equ-26} and \eqref{equ-27}, i.e., $L^a$, after the decoding iteration converges. After obtaining the posterior probability information $L^a$ of the transmitted symbols from the decoder, it is incorporated into the mean and variance updates in Eqs. \eqref{equ-16} and \eqref{equ-17} in Sec. \ref{sec-4-3}.  Consequently, based on Eqs. \eqref{equ-20}-\eqref{equ-21}, the posterior probabilities of the transmission patterns and symbols updated at the observation node $y_{l,m}$ can exploit the symbol posterior information provided by the decoder, thereby yielding more accurate estimation results. Based on the \ac{MAP} criterion, after updating the \ac{LLR} values of the patterns in $\hat{\mathcal{I}}_k^o$, the pattern associated with the maximum \ac{LLR} value is selected as the final detection result, i.e.,
\begin{equation}
	\hat{i}_k^o = \arg \max_j L(b_{k,j}), j\in\hat{\mathcal{I}}_k^o. \label{equ-32}
\end{equation}

 \begin{algorithm} [htpb]
	\setstretch{1.05}
	\caption{ The Joint Pattern Detection and Data Decoding Algorithm}
	\label{alg-2}  
	\begin{algorithmic}[1]	
		\STATE {{\bf Input}: $\mathbf{Y}_c$, $\mathbf{C}$, $\hat{\mathbf{H}}$, $\sigma_n^2$, the candidate pattern set $\hat{\mathcal{I}}_k^o = \{\hat{i}_{k,j}^o, j\in [N_d]\}, k\in \hat{\mathcal{K}}_a$.}\\
		\STATE {{\bf Output}: { The pattern $\hat{i}_k^o$ and data $\hat{\mathbf{u}}_k$, $k \in \hat{\mathcal{K}}_a$}.}\\
		\STATE{{\bf Initial}: The posterior probability $P(x^c_{k,l}=1)=0.5$.}\\

		\% {\sc Joint Pattern Detection and Data Decoding} \\
		\FOR{$k=1$ to $\hat{K}_a$ }
			\FOR{$j=1$ to $N_d$}
				\STATE{ $\hat{i}_{k}^o=\hat{\mathcal{I}}_k^o(j)$, $\hat{i}_{k'}^o = \hat{\mathcal{I}}_{k'}^o(1)$, $k'\in \hat{\mathcal{K}}_a \backslash k$, $p\triangleq\hat{i}_{k}^o$.}\\
				\FOR {$\text{iter}=1$ to $t_{{\rm max}}$}
					\STATE{Calculate extrinsic message $L^e_{l,m \rightarrow k,n}$ via Eq. \eqref{equ-27}.}\\
					\STATE{Calculate posterior message $L^a_{k,n \rightarrow l,m}$ via Eq. \eqref{equ-26}.}\\
					\STATE{Calculate $P(y_{l,m} | h_{k,m}, b_{k,p}, x_{k,l}^c)$ via Eq. \eqref{equ-19}.}\\
					\STATE{Calculate pattern \ac{LLR} $L(b_{k,p})$ via Eq. \eqref{equ-20}.}\\
				\ENDFOR
			\ENDFOR
			\STATE{MAP estimation: $ \hat{i}_k^o = \arg \max_{p} L(b_{k,p}), \quad p \in \hat{\mathcal{I}}_k^o$.}\\
		\ENDFOR \\
		\% {\sc Iterative Polar Decoding Based On Deterministic Patterns}\\
		\FOR{$\text{iter}=1$ to $t_{{\rm max}}$}
			\STATE{Calculate extrinsic message $L^e_{l,m \rightarrow k,n}$ via Eq. \eqref{equ-27}.}\\
			\STATE{Calculate posterior message $L^a_{k,n \rightarrow l,m}$ via Eq. \eqref{equ-26}.}\\
			\STATE{Hard Decision:
        \begin{equation}
            \begin{array}{llll}
                \hat{u}_{k,n} & = & \left\lbrace \begin{array}{ll} 0, & \sum\nolimits_{l,m} L^a_{k,n \rightarrow l,m} > 0 \\ 
                1, & \text{otherwise} \end{array}\right.
            \end{array}, k\in \hat{\mathcal{K}}_a.   \notag
		\end{equation}}\\
		\IF{the CRC check is satisfied for all users}
			\STATE{break;}\\
		\ENDIF
		\ENDFOR
		\STATE{Obtain $\hat{\mathbf{u}}_k$ from $\hat{\mathbf{c}}_k$ via polar transform.}
	\end{algorithmic}  
\end{algorithm}

\par It is worth noting that, in order to improve both pattern detection and data decoding performance, the interaction between data decoding and pattern detection can be iterated multiple times. However, for the consideration of computational complexity and based on simulation results, it suffices to update the posterior \ac{LLR} of transmission patterns using the posterior information from the decoder just only once, which has already achieved satisfactory performance. The schematic of the information exchange process for the above algorithm is presented in Fig. \ref{pic-3}, and the algorithm is summarized in Alg. \ref{alg-2}. Notably, the proposed algorithm performs iterative updates between extrinsic and posteriori information without introducing additional computational overhead. In other words, it only increases the number of iterations without raising the complexity order, and the corresponding complexity analysis will be presented in the next subsection. It is worth noting that the proposed algorithm in this paper is developed for multi-antenna fading scenarios, and it is also applicable to single-antenna fading scenarios, since the latter can be regarded as a special case of the multi-antenna scenario with the number of antennas $M = 1$.

\subsection{Complexity Analysis} \label{sec-4-5}

\par In this subsection, we provide a qualitative analysis of the computational complexity of each algorithm per iteration, measured in terms of the order of floating-point multiplications. The \ac{MMV}-\ac{AMP} algorithm introduced in Sec. \ref{sec-4-1} exhibits a complexity of $\mathcal{O}(n_1N_pM)$ \cite{Liu2018tsp}. For the proposed pattern detection algorithm, the dominant computational cost arises from Eq. \eqref{equ-20}, which is in the order of $\mathcal{O}(K_aN_cn_2M)$. In the \ac{MP}-based polar decoding algorithm, the primary complexity originates from polar decoding, which scales as $\mathcal{O}(K_an_2L_{\rm list}\log n_2)$ since it involves no operation with complexity higher than \ac{SCL} decoding, and $L_{\rm list}$ denotes the list size. Finally, the proposed joint pattern detection and data decoding algorithm maintains the same order of complexity as the algorithms in Sec. \ref{sec-4-2} and \ref{sec-4-3}, as it introduces no additional computational overhead.

\section{Numerical Results} \label{sec-5}

\par In this section, we conduct numerical simulations to rigorously validate the efficacy of the proposed algorithm in both pattern detection and data decoding. The system performance is assessed w.r.t. transmission energy, the number of active users, and the number of received antennas. First of all, we define the parameter settings involved in the simulation, as described in the following subsection. 

\subsection{Parameter Settings}

\par We first specify the iteration numbers for the considered algorithms. The \ac{MMV}-\ac{AMP} algorithm is executed for $30$ iterations. In Algs. \ref{alg-1} and \ref{alg-2}, the number of iterations is set to $t_{{\rm max}}=5$. The numbers of candidate patterns are configured as $N_c=10$ in Sec. \ref{sec-4-3} and $N_d=5$ in Sec. \ref{sec-4-4}. The remaining simulation parameters are detailed in Tab. \ref{tab-1}. Unless otherwise specified, the parameter values listed in this table are adopted throughout the subsequent simulations. To ensure comparability with the baseline scheme FASURA \cite{Gkag2023TCOM}, a relatively large list decoding size, i.e., $L_{\rm list} = 64$, is employed for polar decoding in the simulations; however, in practical systems, the list size can be appropriately reduced to mitigate receiver complexity. In the following, we provide a detailed simulation analysis of the pattern detection and data decoding performance.

\begin{table}[htpb]
	\caption{Parameter settings}
	\label{tab-1}
	\centering
		\begin{tabular}{llll} \toprule
            Parameter & Value & Parameter & Value \\\midrule
             $B_p$ & $16$ & $B_o$ & $16$\\
            $B_c$ & $68$ & $L_p$ & $200$ \\
             $L_c$ & $3000$ & $L$ & $3200$ \\
            $B_{\rm CRC}$ &  $16$ & $n_c$ & $256$\\
		\bottomrule
		\end{tabular}
  \end{table}

\subsection{Pattern Detection Performance}

\par This subsection evaluates the transmission pattern detection performance of three algorithmic schemes: the correlation-based detection in Eq. \eqref{equ-13}, the \ac{MP}-based detection in Eq. \eqref{equ-23}, and the joint pattern detection and data decoding in Sec. \ref{sec-4-4}. For clarity, these schemes are denoted as “Correlation”, “MP-\ac{LLR}”, and “Polar-\ac{LLR}” in the subsequent simulations. In addition, the evaluation of pattern detection performance follows a definition analogous to Eqs. \eqref{equ-3} and \eqref{equ-4}, where the missed detection and false alarm probabilities of the user transmission pattern are denoted by $P_{\rm md}^o$ and $P_{\rm fa}^o$, respectively. Similarly, the overall error probability is denoted by $P_{e}^o = P_{\rm md}^o + P_{\rm fa}^o$. 

\par
\begin{figure*}[t] 
	\begin{minipage}{0.32\linewidth}
		\centerline{\includegraphics[width=\textwidth]{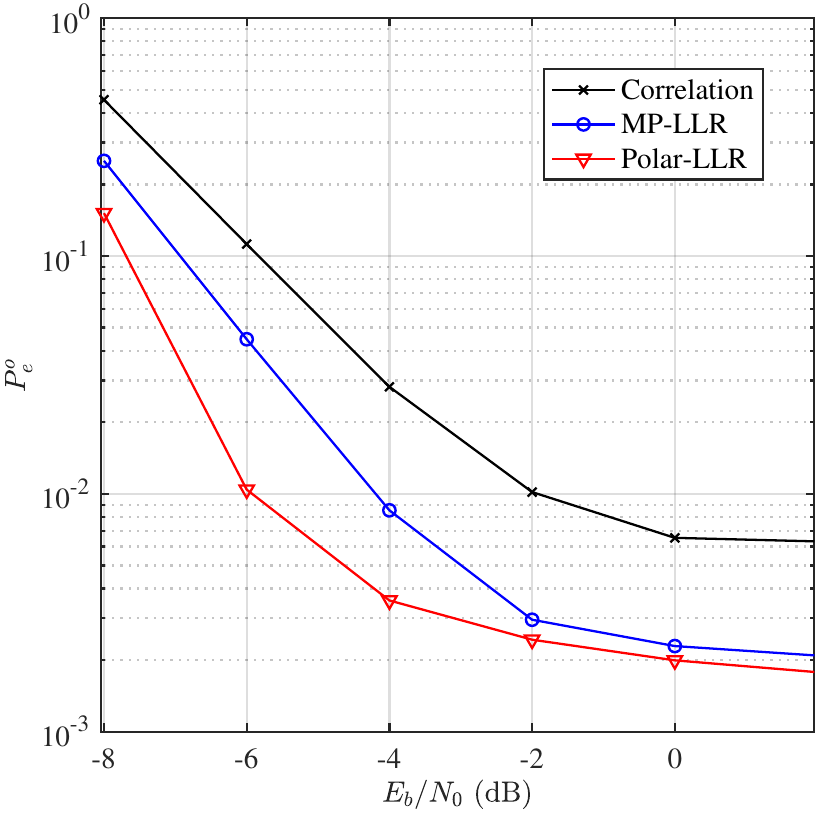}}
		\centerline{(a)}
	\end{minipage}
	\begin{minipage}{0.32\linewidth}
		\centerline{\includegraphics[width=\textwidth]{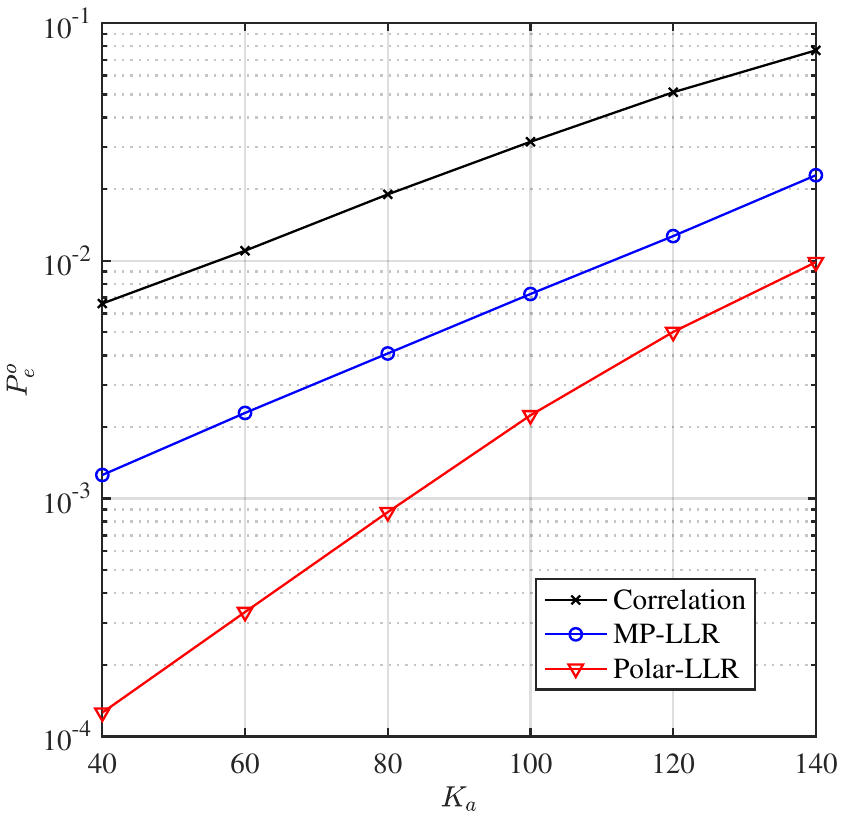}}
		\centerline{(b)}
	\end{minipage}
	\begin{minipage}{0.32\linewidth}
		\centerline{\includegraphics[width=\textwidth]{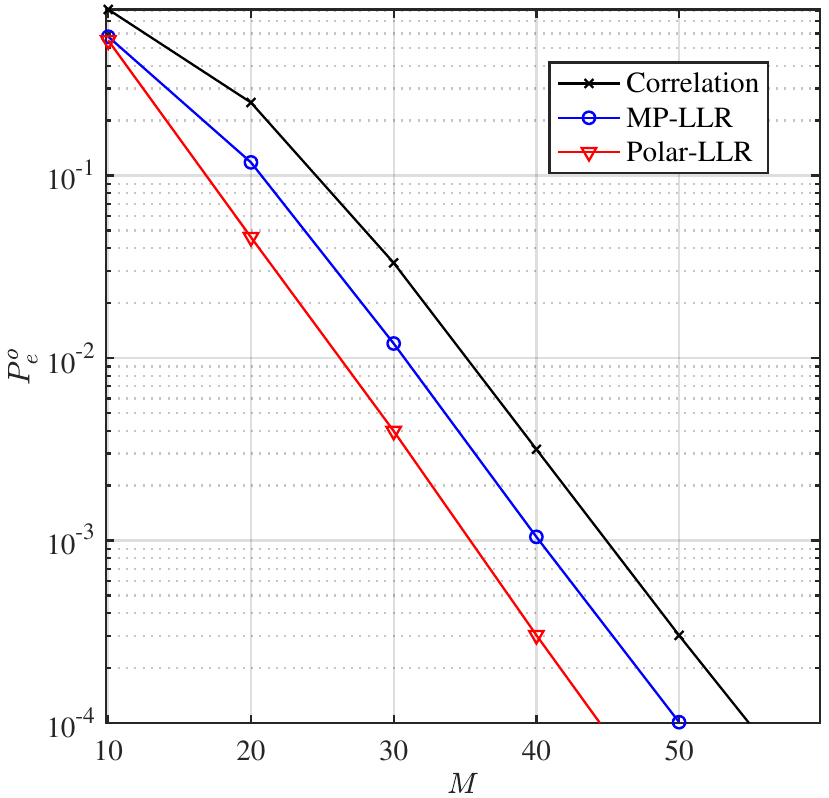}}
		\centerline{(c)}
	\end{minipage}
	\caption{Pattern detection performance evaluation versus (a) $E_b\slash N_0$; (b) $K_a$; (c) $M$, where (a) $K_a=100$, $M=16$; (b) $E_b\slash N_0=-4$ dB, $M=16$; (c) $E_b\slash N_0=-8$ dB, $K_a=100$.}
	\label{pic-simu-1}
\end{figure*}

 Fig. \ref{pic-simu-1} illustrates the pattern detection performance as functions of $E_b\slash N_0$, the number $K_a$ of active users, and the number $M$ of received antennas, respectively. It can be demonstrated in Fig. \ref{pic-simu-1} that, compared with the correlation-based detection, the \ac{MP}-based pattern detection achieves superior detection performance. This improvement stems from the fact that the \ac{MPA} fully exploits the statistical characteristics of the on-off patterns and computes the \ac{LLR}s based on their distribution, thereby enhancing detection accuracy through iterative updates. Furthermore, incorporating the \ac{SO} information of polar codes into the iterative update refines the estimation of symbol's \ac{APP}s, thereby improving the accuracy of conditional \ac{LLR}s and enhancing overall detection performance. For instance, in Fig. \ref{pic-simu-1}(a), for a targeted $P_e^o =10^{-2}$, the required $E_b\slash N_0$ for the above three detection algorithms are approximately $-2$ dB, $-4.2$ dB, and $-6$ dB, respectively, demonstrating the superior energy efficiency of the proposed scheme. Note that as $E_b\slash N_0$ increases, the performance gain of the ''Polar-\ac{LLR}" diminishes. This is primarily because, at higher transmission energy, the \ac{MP}-based detection algorithm already achieves near-optimal performance, while the presence of multi-user interference introduces a performance bottleneck.

\par In Fig. \ref{pic-simu-1}(b), to accommodate different numbers of active users, the codeword length is set to $L_p = 200$ when $K_a \leq 100$, and increased to $240$ and $280$ for $K_a = 120$ and $140$, respectively. The same parameter setting is also applied in Fig. \ref{simu-4}. Generally, a similar conclusion can be drawn, where the proposed pattern detection scheme incorporating the \ac{SO} information of polar codes achieves the best performance. Furthermore, to assess the scalability of the proposed algorithms under large-scale antenna deployments, Fig. \ref{pic-simu-1}(c) presents the pattern detection performance as a function of the number of received antennas, $M$. The results show that the pattern detection algorithm enhanced with the \ac{SO} information of polar decoding achieves the best performance. Notably, increasing $M$ does not lead to any evident performance bottleneck across the detection schemes. This suggests that in multi-user transmission scenarios, when further increases in $E_b\slash N_0$ fail to improve system performance, augmenting the number of received antennas can leverage spatial diversity to enhance multi-user access capabilities.

\subsection{Data Decoding Performance}

\par In this subsection, we assess the efficacy of the proposed scheme in data decoding and benchmark its performance against the following approaches:
\begin{itemize}
	\item \textbf{\textit{IDMA-LDPC}}: The sparse \ac{IDMA} scheme using \ac{LDPC} codes \cite{li2022JSAC}, which adopts a two-stage encoding approach, with $B_p=16$, $B_c=84$, $L_p=200$, and $L_c=3000$.
	\item { \textbf{\textit{ODMA-LDPC}}: The \ac{ODMA} scheme using \ac{LDPC} codes \cite{li2022JSAC}, where the coding parameters are align with Tab. \ref{tab-1}.}
	\item \textbf{\textit{IDMA-Polar}}: The sparse \ac{IDMA} scheme using polar codes, where the parameters are aligned with \textit{IDMA-LDPC}, and the \ac{MP}-Based \ac{SO} polar decoding is utilized.
	\item \textbf{\textit{ODMA-Polar}}: The proposed scheme in this paper, where the \ac{JDD} algorithm introduced in Sec. \ref{sec-4-4} is employed.
	\item \textbf{\textit{Spread-Polar}}: The \ac{SO} polar-coding scheme based on symbol spreading, with the polar-coded length $256$, the spreading length $11$, and $B_p=16$, $B_c=84$, $n_1=384$.
	\item \textbf{\textit{FASURA}}: The spreading scheme based on \ac{HO} polar codes \cite{Gkag2023TCOM}, which acts as the \ac{PUPE} benchmark in the \ac{MIMO}-\ac{URA} scenario.
	\item \textbf{\textit{Coupled-ODMA}}: The \ac{ODMA} scheme employing the \ac{HO}-\ac{SCL} polar decoding algorithm \cite{Ozates2024GC}.
\end{itemize}

\par
\begin{figure}[htpb]
	\centerline{\includegraphics[width=0.45\textwidth]{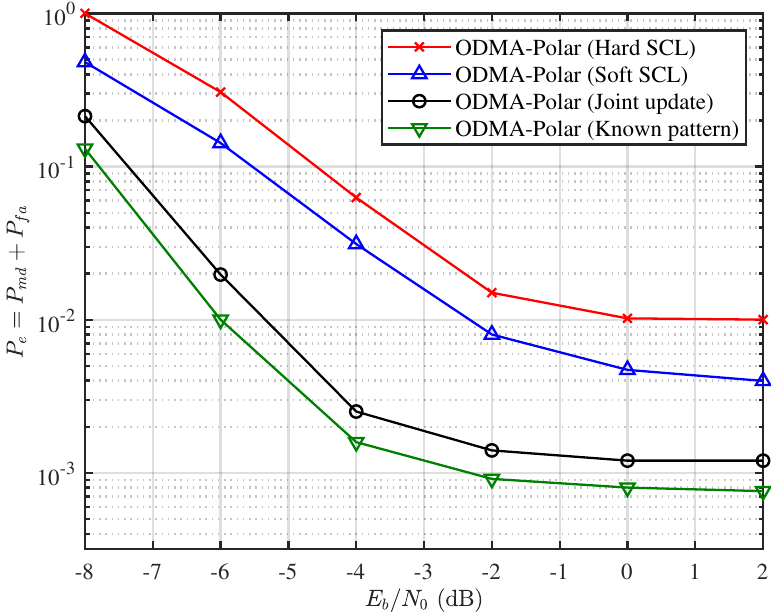}}
	\caption{The comparison of polar decoding performance under different schemes versus $E_b \slash N_0$, where $K_a = 100$ and $M = 16$.}
	\label{simu-2}
\end{figure}

 Firstly, we assess the performance gains of the proposed \ac{MP}-based \ac{SO}  polar decoding algorithm under different conditions, as illustrated in Fig. \ref{simu-2}. The legends “Hard SCL”, “Soft SCL”, “Joint SCL”, and “Known Pattern” correspond to the conventional \ac{HO} \ac{SCL} polar decoding scheme \cite{SCL}, the proposed \ac{SO} decoding scheme introduced in Sec. \ref{sec-4-3}, the joint pattern detection and polar decoding scheme presented in Sec. \ref{sec-4-4}, and the ideal case where the on-off patterns are perfectly known, respectively. The results in Fig. \ref{simu-2} demonstrate that, compared with the \ac{HO} polar decoding algorithm, exploiting the \ac{SO} information of polar codes and performing \ac{MP}-based \ac{JDD} yields a substantial performance improvement. Moreover, incorporating the \ac{SO} information into the iterative detection of on-off patterns not only increases the detection accuracy but also further boosts data decoding performance. Based on these observations, the “Joint SCL” strategy is adopted as the default decoding method for the ODMA-Polar receiver in the subsequent simulations.

\par
\begin{figure}[htpb]
	\centerline{\includegraphics[width=0.45\textwidth]{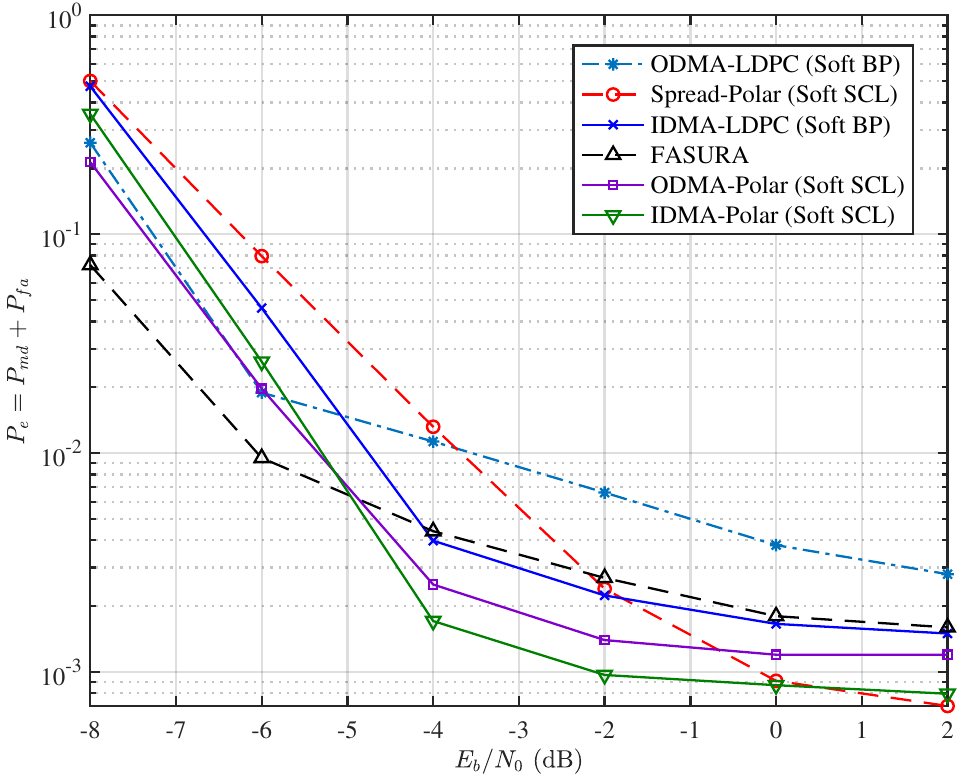}}
	\caption{ The performance comparison of \ac{URA} schemes versus $E_b \slash N_0$, where $K_a = 100$ and $M = 16$.}
	\label{simu-3}
\end{figure}

 Furthermore, Fig. \ref{simu-3} illustrates the decoding performance of different \ac{URA} schemes as a function of $E_b/N_0$. The results indicate that all schemes reach a performance plateau as $E_b/N_0$ increases. In Region 1 ($E_b/N_0 \in [-8\ \text{dB}, -5\ \text{dB}]$), FASURA \cite{Gkag2023TCOM} attains the best decoding performance, whereas in Region 2 ($E_b/N_0 \in [-5\ \text{dB}, 2\ \text{dB}]$), IDMA-Polar outperforms the others. In contrast, the ODMA-based polar decoding scheme slightly outperforms IDMA-Polar in Region 1 but falls slightly behind in Region 2. This is mainly because, although ODMA reduces the polar code rate compared with IDMA—potentially enhancing decoding performance—its performance is limited by the accuracy of pattern detection. While this characteristic is also observed in the \ac{IDMA}-\ac{LDPC} and \ac{ODMA}-\ac{LDPC} schemes. Moreover, both IDMA- and ODMA-based polar schemes outperform the LDPC-based scheme and the symbol-spreading polar scheme, indicating that polar codes achieve superior performance under short blocklengths and that sparse-interleaving multiple access architectures (IDMA/ODMA) offer advantages over symbol-spreading schemes.

\par
\begin{figure}[htpb]
	\centerline{\includegraphics[width=0.45\textwidth]{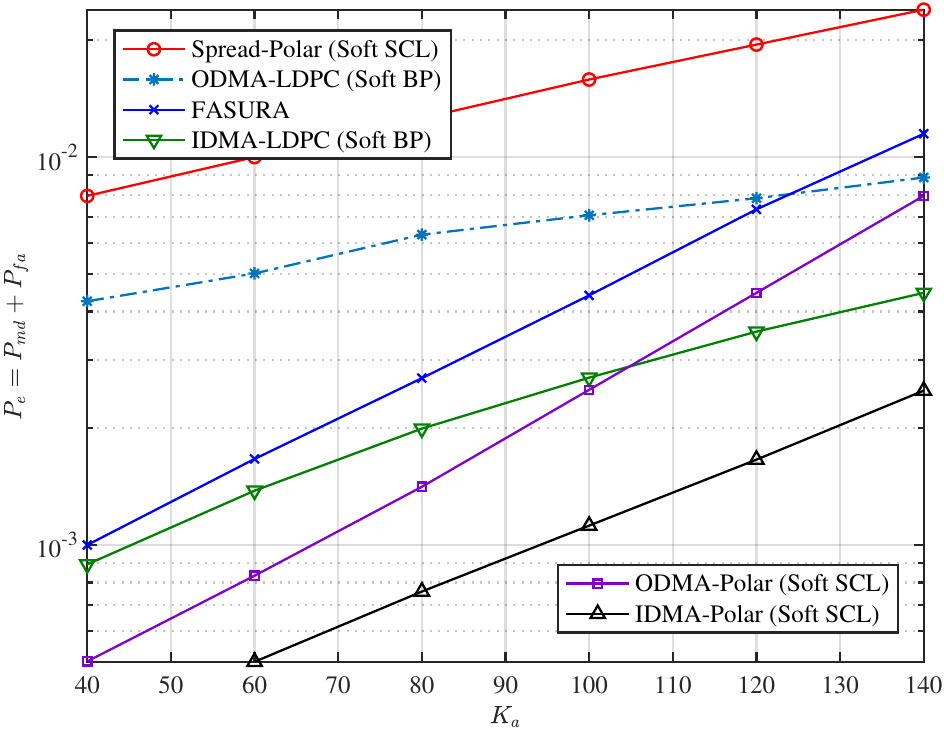}}
	\caption{ The performance comparison of \ac{URA} schemes versus $K_a$, where $E_b\slash N_0 = -4$ dB and $M = 16$.}
	\label{simu-4}
\end{figure}

 Additionally, Fig. \ref{simu-4} shows the decoding performance of different \ac{URA} schemes as a function of $K_a$. Overall, the IDMA/ODMA-based architecture outperforms the symbol-spreading FASURA scheme while the \ac{ODMA}-\ac{LDPC} scheme is a little bit worse. However, the performance of the \ac{ODMA}-based polar decoding algorithm is constrained by the accuracy of pattern detection, leading to a slight degradation relative to the \ac{IDMA}-\ac{LDPC} scheme under high user loads. In contrast, when integrated with joint \ac{MIMO} detection, polar codes within the IDMA framework exhibit superior decoding performance compared to LDPC codes.

\par
\begin{figure}[htpb]
	\centerline{\includegraphics[width=0.45\textwidth]{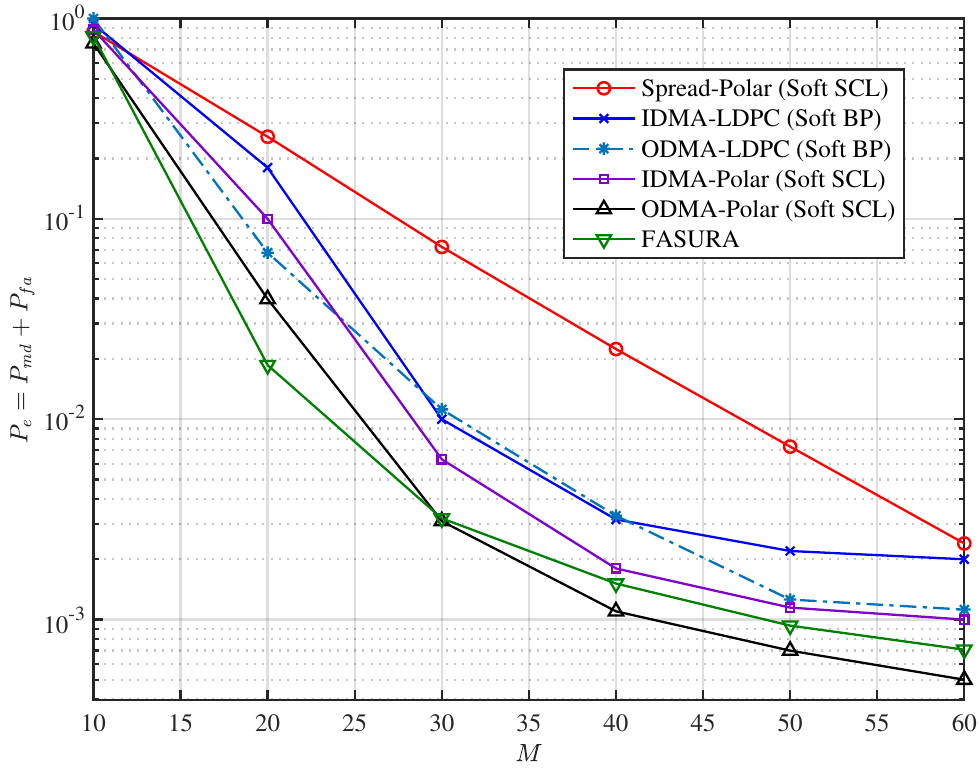}}
	\caption{ The performance comparison of \ac{URA} schemes versus $M$, where $E_b\slash N_0 = -8$ dB and $K_a = 100$.}
	\label{simu-5}
\end{figure}

 To further investigate the potential performance benefits of the proposed \ac{ODMA}-Polar scheme, we evaluate the performance of various \ac{URA} schemes as a function of $M$, as shown in Fig. \ref{simu-5}. The results indicate that when $M \geq 30$, the ODMA-based polar coding scheme outperforms the others, highlighting its advantages in large-scale antenna configurations. This is primarily attributed to the exploitation of multi-antenna diversity gain, which enhances pattern detection accuracy and further improves overall decoding performance. This trend is also evident in Fig. \ref{pic-simu-1}(c). With large-scale antenna configurations, the improved \ac{ODMA} pattern detection accuracy, reduced polar coding rate, and superior performance of the \ac{MP}-based \ac{SO} polar decoding algorithm collectively contribute to the significant enhancement of the proposed \ac{ODMA}-Polar scheme, demonstrating superior detection accuracy and decoding performance. Similarly, when the number of antennas exceeds $40$, the \ac{ODMA}-\ac{LDPC} scheme also outperforms the \ac{IDMA}-\ac{LDPC} scheme. This further indicates that, as the number of antennas increases, reliable transmission pattern detection can still be maintained under the \ac{ODMA} architecture, thereby demonstrating its advantage over the \ac{IDMA} architecture.

\par
\begin{figure}[htpb]
	\centerline{\includegraphics[width=0.45\textwidth]{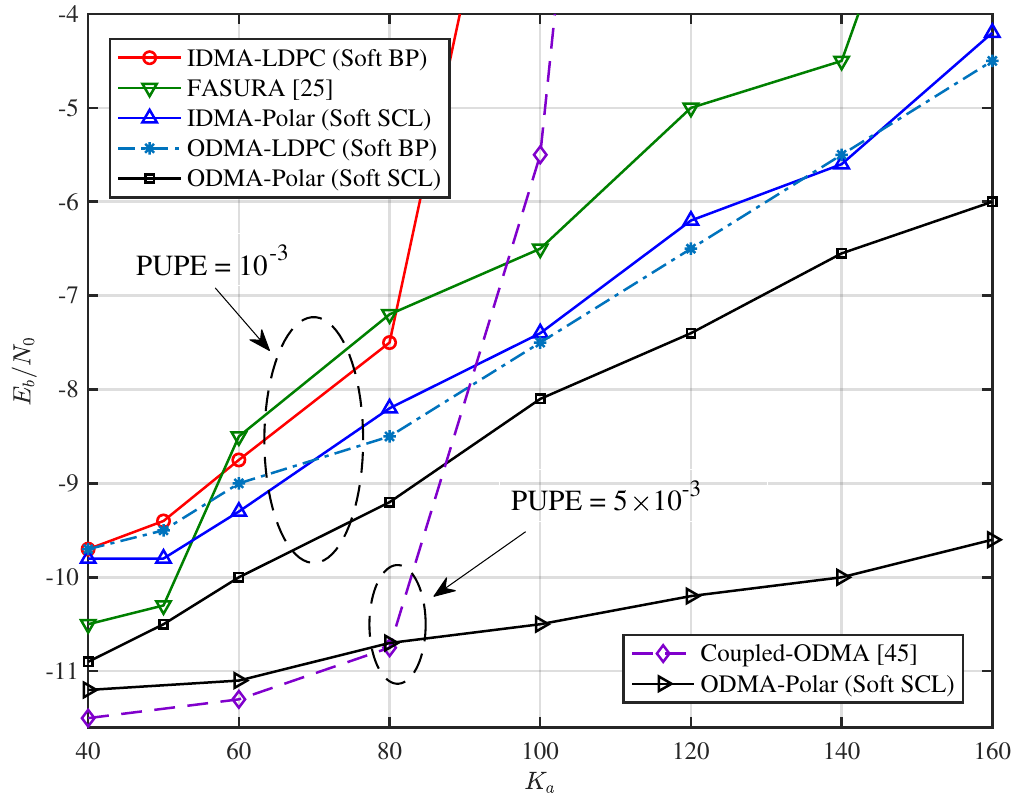}}
	\caption{The required $E_b\slash N_0$ for different schemes versus $K_a$ under the target \ac{PUPE} $P_e=5\times10^{-3}$ for Coupled-\ac{ODMA} scheme \cite{Ozates2024GC} and $P_e = 10^{-3}$ for other schemes. The number of receiving antennas is set to $M=50$. }
	\label{simu-6}
\end{figure}

 Finally, we evaluate the minimum $E_b/N_0$ required to achieve a given \ac{PUPE} as a function of $K_a$, as shown in Fig. \ref{simu-6}. It is worth noting from the simulation results in Figs. \ref{simu-3}-\ref{simu-5} that, although {FASURA} achieves better performance in the \ac{PUPE} regime of $P_e = 10^{-2}$, its performance is gradually surpassed by the \ac{IDMA}/\ac{ODMA}-based schemes as the \ac{PUPE} decreases further. This indicates that the proposed schemes offer significant advantages in mitigating multi-user interference and reducing the error floor. To further demonstrate the superiority of the proposed algorithms in handling multi-user interference, the target \ac{PUPE} regime is set to $P_e = 10^{-3}$. On such basis, with the number of antennas fixed at $M=50$, we further evaluate the minimum $E_b\slash N_0$ required by different algorithms to achieve the target \ac{PUPE}. The considered schemes include: \ac{IDMA}-\ac{LDPC}, \ac{ODMA}-\ac{LDPC}, \ac{IDMA}-Polar, and \ac{ODMA}-Polar, with FASURA \cite{Gkag2023TCOM} and \ac{IDMA}-LDPC \cite{Pradhan2022TCOM} serving as the benchmarks.

\par It is worth noting that \cite{Zhang2024TIOJ,Zhang2025TVT} also investigated receiver designs for pilot-uncoupled \ac{ODMA} schemes in \ac{MIMO} channels. However, the algorithm proposed in \cite{Zhang2024TIOJ} relies on high-complexity matrix decomposition operations and is therefore unsuitable for heavily loaded scenarios (i.e., $K_a > 30$), making it impractical as a benchmark for comparison. As for \cite{Zhang2025TVT}, the considered scheme is essentially based on a slotted transmission architecture, where slot selection additionally conveys information bits, thus preventing a fair comparison with the considered schemes. In addition, we evaluate the performance of the \ac{HO} polar decoding algorithm under the pilot-coupled \ac{ODMA} framework proposed in \cite{Ozates2024GC}. According to the reported results, the scheme achieves excellent performance in the \ac{PUPE} regimes of $P_e=0.05$ and $P_e=0.1$. However, our simulations show that its performance remains constrained by multi-user interference, preventing reliable operation in the lower-\ac{PUPE} regime of $P_e=10^{-3}$. Nevertheless, to further illustrate the insights provided by this scheme, we additionally evaluate its performance for several $K_a$ settings in the \ac{PUPE} regime of $P_e=0.005$, as shown in Fig. \ref{simu-6}.

\par As depicted in Fig.~\ref{simu-6}, the performance advantage of the proposed \ac{ODMA}-Polar scheme is evident. This advantage arises from two factors: First, the coding gain from the superior decoding performance of polar codes under short blocklengths, along with the proposed joint pattern detection and data decoding architecture based on the soft information of polar codes; second, the performance gain from the \ac{ODMA} framework, which outperforms the \ac{IDMA} architecture in large antenna configurations. As a result, the final findings validate that the \ac{ODMA}-Polar scheme outperforms the other two schemes, with an average gain of approximately $1$ dB relative to the \ac{IDMA}-Polar scheme and more than $2$ dB gain to FASURA at the target PUPE of $P_e=10^{-3}$. Moreover, for a target PUPE of $P_e=5\times 10^{-3}$, the coupled-\ac{ODMA} scheme \cite{Ozates2024GC} outperforms the proposed \ac{ODMA}-Polar scheme in the regime where $K_a <80$ but ceases to function when $K_a>80$. In contrast, the \ac{ODMA}-Polar scheme maintains robust performance across the entire range of $K_a$ from $40$ to $160$. Notably, it continues to operate reliably under low $E_b\slash N_0$ conditions even when $K_a$ exceeds $80$, further corroborating its superior capability in mitigating multi-user interference and suppressing the \ac{PUPE} floor.

\section{Conclusion} \label{sec-6}

\par In this paper, we proposed an \ac{ODMA}-based transmission scheme for \ac{MIMO} massive {URA} systems, incorporating \ac{SO} polar codes. First, we introduced the pilot-uncoupled three-segment coding scheme within the \ac{ODMA} framework. The \ac{MMV}-\ac{AMP} algorithm was then employed for effective active user detection and channel estimation. Subsequently, We proposed a hierarchical pattern detection framework to improve detection accuracy while reducing computational complexity. Specifically, coarse-grained pattern estimation was first obtained via correlation operations, followed by iterative posterior probability updates through a message-passing algorithm and subsequent \ac{MAP}-based pattern detection. Furthermore, a joint pattern detection and data decoding algorithm was developed by exploiting the \ac{SO} posterior probability information from polar decoding to further enhance both pattern detection and decoding performance. Simulation results demonstrated that the proposed scheme achieved significant performance improvements over existing \ac{URA} techniques, particularly under large-scale antenna configurations.

\section{Acknowledgement}

\par We gratefully acknowledge Dr. Ozates \cite{Ozates2024GC} and Dr. Zhang \cite{Zhang2025TVT} for providing simulation results on \ac{ODMA}-\ac{URA} works, and Prof. Yuan \cite{Yuan2025TIT} for providing simulation results on \ac{SO}-\ac{SCL} polar decoding.

\end{document}